\documentclass[aip,reprint,
							floatfix,
						 	twocolumn,
							superscriptaddress,
				            notitlepage,
							footinbib,
							floatfix,
						longbibliography]{revtex4-2}

\usepackage{graphicx}
\usepackage{dcolumn}
\usepackage{bm}		
\usepackage{amsmath}
\usepackage{amssymb}
\usepackage{mathbbol}

\usepackage[utf8]{inputenc}

\usepackage[normalem]{ulem}

\usepackage[usenames,dvipsnames]{xcolor}
\usepackage{slashed}

\usepackage{microtype}

\usepackage{rotating}					
\usepackage{todonotes}

\DeclareSymbolFont{rsfs}{U}{rsfs}{m}{n}
\DeclareSymbolFontAlphabet{\mathrsfs}{rsfs}
\usepackage[mathscr]{eucal}

\usepackage{hyperref}
\hypersetup{colorlinks=true,citecolor=MidnightBlue,urlcolor=BrickRed,linkcolor=Green}

\usepackage{booktabs}

\let\ud\d
\let\vec\boldsymbol

\newcommand{\Ai}{\mathrm{Ai}\,}

\newcommand{\deriv}[2]{\frac{\mathrm d #1}{\mathrm d #2}}

\newcommand{\uvb}{\hat{\mathcal B}}
\newcommand{\uve}{\hat{\mathcal E}}
\newcommand{\uvk}{\hat{\mathcal K}}

\begin{document}

\title{Kinetic theory for spin-polarized relativistic plasmas}

\author{Daniel Seipt}
\email{d.seipt@hi-jena.gsi.de}
\affiliation{Helmholtz Institut Jena, Fr\"obelstieg~3, 07743 Jena, Germany}
\affiliation{GSI Helmholtzzentrum für Schwerionenforschung GmbH, Planckstrasse 1, 64291 Darmstadt, Germany}

\author{Alec G. R. Thomas}
\affiliation{The G\'erard Mourou Center for Ultrafast Optical Science, University of Michigan, Ann Arbor, Michigan 48109, USA}

\begin{abstract}
{The investigation of spin and polarization effects in ultra-high intensity laser-plasma and laser-beam interactions has become an emergent topic in high-field science recently. In this paper we derive a {relativistic} kinetic description of spin-polarized plasmas, where QED effects are taken into account via Boltzmann-type collision operators {under the local constant field approximation}. The emergence of anomalous precession is derived from one-loop self-energy contributions in a strong background field. We are interested, in particular, in the interplay between radiation reaction effects and the spin polarization of the radiating particles. For this we derive equations for spin-polarized quantum radiation reaction from moments of the spin-polarized kinetic equations. By comparing with the classical theory, we identify and discuss the spin-dependent radiation reaction terms, and radiative contributions to spin dynamics.}
\end{abstract}

\keywords{Spin polarization, kinetic equation, radiation reaction}

\maketitle

\section{Introduction}

There has recently been substantial recent interest in the effects of lepton and gamma-ray spin/polarization on strong-field quantum-electrodynamics (QED) emission processes that occur in ultraintense laser interactions with electron beams or plasma \cite{Ivanov_QED_spin, Ivanov_QED_spin2,King_PRA_2013, Sorbo_PRA, Seipt_PRA_2018,Seipt_PRA_2019, Sorbo_PPCF, Seipt_PRA_2020, Seipt_NJP_2021, Chen_PRL_2019, Li_PRL_2019, Li_PRL_2020_gamma_ray, Li_PRL_2020_helicity_transfer, Wan_PRR_2020, guo_stochasticity_2020, Thomas:2020, buscher_generation_2020, Dai_MRE_2022, king_nonlinear_2020}. This is motivated by the continuing development of multi-petawatt class laser systems around the world, which should enable researchers to access QED critical strength fields in combination  with relativistic plasma dynamics \cite{Zhang:Perspective}. Strong field QED is important for particles with momentum $p^\mu$ interacting in electromagnetic fields for which the invariant quantity $\chi_p = ||F^{\mu\nu}p_\nu||/{(m c E_{S})} \equiv  {\left[\gamma\sqrt{(\vec{E}+\vec{\beta}\times \vec{B})^2-(\vec{\beta}\cdot\vec{E})^2}\right]/E_{S}}$ is of order 1 or larger, where $E_{S} = m^2c^3/|e|\hbar$ is the Sauter-Schwinger field, often also called the QED critical field. {The {electromagnetic} fields in the laboratory frame are typically much weaker than $E_S$.} Under such conditions, the {dominant quantum processes are the} first order processes of photon emission by a lepton (nonlinear Compton scattering, in a laser field) or the decay of a photon into an electron-positron pair (the nonlinear Breit-Wheeler process). These processes depend on the spin/polarization state of both the leptons and photons \cite{Ivanov_QED_spin, Ivanov_QED_spin2,Seipt_PRA_2020,chen_electron_2022,Torgrimsson:NJP2021}. 

Studies of strong field QED have often made use of a semiclassical approach where the particles follow classical trajectories punctuated by point-like quantum emission/decay events that are calculated probabilistically under the assumptions that the `formation length' is short compared to the characteristic space/time scales of the electromagnetic fields (the local constant field approximation, LCFA);  only first order processes are dominant; the fields are ``weak'' such that the field configuration in the rest frame of the particle is a crossed electric and magnetic field \cite{Ridgers:JCompPhys2014,Blackburn:RevModPlasma2020,Gonoskov_RMP_2022}. Motivated by this, recent studies have used an extended version of this model where the emission events are modified to include polarization dependent quantum rates combined with a classical spin-pusher using the Thomas Bargmann-Michel-Telegdi (T-BMT) equation \cite{Sorbo_PRA, Seipt_PRA_2018,Seipt_PRA_2019, Sorbo_PPCF, Seipt_PRA_2020, Seipt_NJP_2021,Chen_PRL_2019, Li_PRL_2019, Li_PRL_2020_helicity_transfer}. However, this is done on an ad-hoc basis, and it is not immediately clear that, for example, the inclusion of the anomalous ($g-2$) magnetic moment in the T-BMT equation is consistent with the other assumptions. Other authors have been interested in the feedback of electron spin effects on classical expressions of radiation reaction \cite{Li_JCP_2021,Wen:PRA2017,geng_spin-dependent_2020,torgrimsson_resummation_2021}. In this paper, we develop a kinetic description for leptons in strong fields starting from transport equations resulting from a Wigner operator formalism in the mean field approximation \cite{Vasak:AnnPhys1987} and include  coupling of the fermion field to the high energy photons through a Boltzmann-like collision operator representing the point-like quantum emission/decay events \cite{Neitz:PRL2013}. By taking moments of a delta-distribution in phase-space we derive {effective quasi-}classical equations of motion for leptons, including the effects of spin and radiation reaction. 

We start in section \ref{sect:classical} by reviewing the ``classical'' equations of motion with radiation reaction and note some properties, including the relative strength of spin corrections. In section \ref{sect:kinetic} we derive the set of Boltzmann-like transport equations for the particles, comprising the classical transport and collision type operators for the quantum emissions. Section \ref{sect:moments} introduces the fluid equations for the transport by taking moments of the distribution functions \cite{Ridgers:JPP2017}. In section \ref{sect:singleparticle}, by assuming a delta-distribution, we derive the single particle equations of motion and discuss the consequences of the expressions. Further discussion of the results is presented in section \ref{sect:discussion}.
We conclude in section \ref{sect:conclusions}. We employ rationalized Heaviside-Lorentz units with $\hbar=c=\varepsilon_0=1$ and Minkowski metric $\eta^{\mu\nu} = {\rm diag}(1,-1,-1,-1)$ throughout. 
We define dimensionless electron momenta according to $p^\mu/m\to p^\mu$ and coordinates according to $mx^\mu \to x^\mu$, and $m\tau\to\tau$ for proper time. We also absorb the magnitude of the electron charge $|e| = \sqrt{4\pi\alpha}$, where $\alpha$ is the fine structure constant, and the electron mass into the  electromagnetic field strength according to
$|e|F^{\mu\nu}/m^2\to F^{\mu\nu}$.
This means that in all the following equations, a factor $q=\pm1$ simply gives the sign of charge.

\section{Classical Radiation Reaction and Spin}
\label{sect:classical}

\subsection{Classical Radiation Reaction for Orbital Motion}
\begin{table*}[t]
\caption{{Summary of  radiation reaction models and their effect on spin precession.}}
\label{tab:RR}
\begin{tabular}{cccc}
\toprule
&  $R^\mu$ & $P^{\mu\nu}R_\nu$ & $S.R$ \\
\midrule
LAD \cite{Dirac:ProcRoySoc1938} & 
		$\ddot u^\mu$ & 
		$\ddot u^\mu + u^\mu \dot u^2 $ 		&
		$S.\ddot u$ \\
EFO \cite{Ford:PLA1993}& 
		$ 
		 q(u.\partial) F^{\mu\nu} u_\nu + q F^{\mu\nu} \dot u_\nu$ & 
		 $ q(u.\partial) F^{\mu\nu} u_\nu + q F^{\mu\nu} \dot u_\nu - q u^\mu u.F.\dot u$ &
		$ q (u.\partial) S. F.u + q S.F.\dot u $ \\
MP \cite{Mo:PRD1971} &  
		$ q F^{\mu\nu} \dot u_\nu $ & 
		$ q F^{\mu\nu} \dot u_\nu - q u^\mu u.F.\dot u$ & 
		$ q S.F.\dot u$ \\
LL \cite{book:Landau2} & 
		$  q (u.\partial) F^{\mu\nu}u_\nu +  F^{\mu\nu}F_{\nu\kappa} u^\kappa $ &
		$ q (u.\partial) F^{\mu\nu} u_\nu 
		+  F^{\mu\nu} F_{\nu\kappa} u^\kappa  
		- u^\mu  \, u . F^2. u $ &
		$ q (u.\partial ) S.F.u +  S.F^2.u  $ \\
H  \cite{Herrera:PRD1977} & 
		$F^{\mu\nu}F_{\nu\kappa} u^\kappa $ & 
		$  F^{\mu\nu} F_{\nu\kappa} u^\kappa  
		- u^\mu  \, u . F^2. u$ &
		$S.F^2.u$ \\
\bottomrule
\end{tabular}
\end{table*}
It is a well known fact that the radiation emitted by accelerated charges acts back onto the electron motion \cite{book:Jackson}. For the classical orbital equations of motion these so-called radiation reaction effects can be taken into account by adding a radiation reaction force $f_\mathrm{rad}^\mu$ to the Lorentz force,
\begin{align} \label{eq:RR}
\deriv{u^\mu }{\tau} = q F^{\mu\nu} u_\nu + {f}_\mathrm{rad}^\mu \,,
\end{align}
{where $\tau$ is the particle's proper time, $u^\mu$ is the four-velocity/momentum and $F^{\mu\nu}$ is the electromagnetic field tensor.}
Many forms of the radiation reaction force {$f_\mathrm{rad}^\mu$} have been proposed in the literature, but no definite consensus
on the ``correct'' equation of motion has been reached so far.
Let us write the radiation reaction force generically as follows:
$ {f}_\mathrm{rad}^\mu  = \tau_R  \: \Delta^{\mu\nu} R_\nu $, with the {dimensionless radiation reaction time constant
$\tau_R = 2\alpha/3 $}, and a projector $ \Delta^{\mu\nu} =  {\eta}^{\mu\nu} - u^\mu u^\nu $ onto the three-dimensional space-like hyperplane perpendicular to the four-velocity $u^\mu$. 
This is necessary to ensure that the radiation reaction force and hence the total acceleration of the particle is orthogonal to its velocity, i.e. to ensure the validity of the subsidiary condition $u^2=1$.

The specific form of the vector $R^\mu$ varies for the different models of classical radiation reaction (RR). In particular, for the Lorentz-Abraham-Dirac (LAD) form of the RR force we have \cite{Lorentz1909,Abraham,Dirac:ProcRoySoc1938}
\begin{align} \label{eq:F-LAD}
R^\mu_\mathrm{LAD} =  \ddot u^\mu \,,
\end{align}
such that ${f}_\mathrm{rad}^\mu = \tau_R (\ddot u^\mu + u^\mu \dot u^2)$ after an integration by parts.

However, the LAD form of radiation reaction is known to have some mathematical issues in the form of runaway solutions and preacceleration. These phenomena are related to the fact that LAD contains the second derivative of $u$, a third initial condition---the initial acceleration---has to be provided. However, it is not independent and only a proper choice leads to physical solutions of the LAD
(see also detailed discussion in Refs.~\cite{Rohrlich:AnnPhys1961,Klepikov:PhysUspekh1985}).

Many alternative forms of the equations of motion for classical radiation reaction have been derived in order to cure the deficiencies of the LAD equation. Most notably, the Landau-Lifshitz (LL) form of radiation reaction has been derived from LAD by reducing the order of the differential equation by iteration \cite{book:Landau2}. (See also the recent Refs.~\cite{ekman_reduction_2021,ekman_reduction_2022}.)
Iteration means treating the radiation reaction term in the LAD equation as a perturbation and approximating the jerk $\ddot u$ {using} the Lorentz force, $\ddot u = q \deriv{(F.u)}{\tau} = q (u.\partial F).u +q F.\dot u$. {Iterating the LAD equation twice with this expression} yields the LL equation, which has attracted some considerable interest in the recent years due to its usefulness in numerical simulations of high-intensity laser-plasma interactions.

In fact, many of the RR models in the literature are connected to LAD by successive iterations and the quasi-constant approximations, even though their original derivation often was not following this path. Iterating LAD only once immediately leads to the Eliezer/Ford-O'Connell (EFO) equation \cite{Ford:PLA1993}, iterating a second time directly gives the Landau-Lifshitz (LL) form of RR as mentioned above.
In addition to iterating the RR equations one can also take a quasi-constant approximation by neglecting the $u.\partial F$ terms. The quasi-constant approximation of EFO is known as Mo-Papas (MP) equation \cite{Mo:PRD1971}, while  the quasi-constant limit of LL is the Herrera (H) equation \cite{Herrera:PRD1977}. 
The specific forms of the radiation reaction vector $R^\mu$ and the RR force $\Delta^{\mu\nu} R_\nu$ are {summarized} in Table~\ref{tab:RR}.
For a discussion of the different RR models in general see \cite{Burton:ContempPhys2014}, and as a limit of QED see also Ref.~\cite{Ilderton:PRD2013b}.

\subsection{Covariant form of spin precession with radiation reaction}
The {well established} relativistic equation of motion for the covariant spin {four}-vector of a particle in a slowly varying external field is the Bargmann-Michel-Telegdi (BMT) equation \cite{Bargmann:PRL1959}, which reads
\begin{align}
\deriv{S^\mu}{\tau} & = {q}\frac{ g}{2} 
\left[ F^{\mu\beta} S_\beta + u^\mu (S_\alpha F^{\alpha\beta} u_\beta ) \right] - u^\mu (S_\alpha \dot u^\alpha )\,,
\label{eq:BMT-dotu}
\end{align}
where $S^\mu$ is the spin four-vector and $g$ is the Landé-factor.  For an electron, $g\approx 2$, and the deviation from 2 is the anomalous magnetic moment $a_e \equiv (g-2)/2$.
The one-loop perturbative QED result derived by Schwinger \cite{Schwinger:PR1951} is $a_e^\mathrm{1-loop}=\alpha/2\pi$, and the experimentally measured value is
$a_e = 1.15965218076(28) \times 10^{-3}$ \cite{Workman:2022ynf}.

The last term in Eq.~\eqref{eq:BMT-dotu} ensures that the spin four-vector $S^\mu$ remains orthogonal to the four-velocity $u^\mu$. This is necessary since the spin-four vector is space-like, and in the electron rest frame has no time-component. The form of Eq.~\eqref{eq:BMT-dotu} ensures that $S.u=0$ as long as $u^2=1$, which then ensures that the length of the spin four-vector is conserved under the BMT equation $\deriv{(S.S)}{\tau} =0$. This now has the consequence that the T-BMT equation also acquires a contribution from radiation reaction. Taking the form for the radiation reaction force as described above and plugging 
this into \eqref{eq:BMT-dotu} we get
\begin{align}
\deriv{S^\mu}{\tau} = {q}\frac{g}{2} F^{\mu\nu}S_\nu - q a_e\, u^\mu\: (u.F.S) - \tau_R \, u^\mu \: (S.R)  \,,
\end{align}
where the last term appears due to the radiation reaction force. Clearly, different forms of $R^\mu$ for the various RR models will yield different contributions to the T-BMT equation, {as shown in the} last column of Table~\ref{tab:RR}. 

\begin{figure}[ht]
    \centering
    \includegraphics[width=\columnwidth]{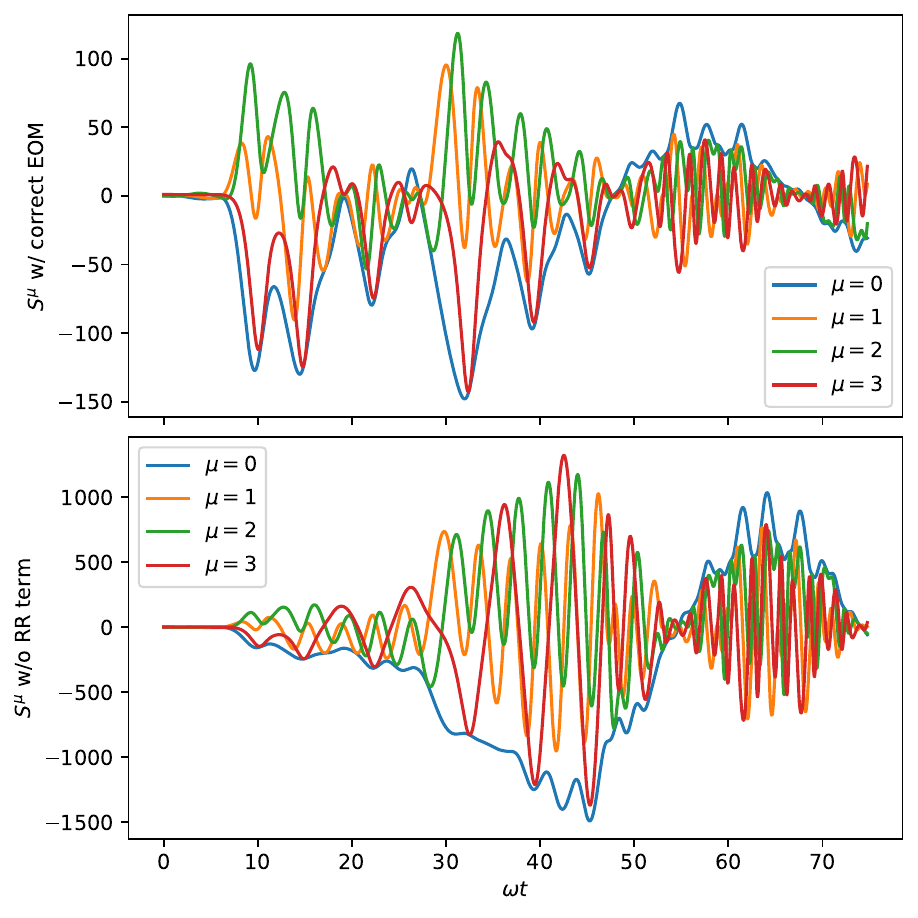}
    \caption{Evolution of the spin-vector $S^\mu$ in a linearly polarized standing wave with $a_0=200$, obtained by numerically solving Eqns.~\eqref{eq:orbcor} and \eqref{eqn:spincor}. In the lower panel the RR term in the BMT equation has been omitted, resulting in a completely wrong behaviour of $S^\mu$.}
    \label{fig:SR}
\end{figure}

Employing for instance the Herrera RR-force (LL in a quasi-constant field), we immediately get as the most simple consistent classical RR model with spin-precession:
\begin{align} \label{eq:orbcor}
\deriv{u^\mu}{\tau} 
		&=
			q F^{\mu\nu} u_\nu + \tau_R \, \left[ 
			F^{\mu\nu} F_{\nu\lambda} u^\lambda
				- u^\mu ( u_\alpha F^{\alpha\beta} F_{\beta\lambda} u^\lambda )
			\right] \\
\deriv{S^\mu}{\tau} 
		& = q\frac{g}{2} F^{\mu\nu} S_\nu
			- q a_e u^\mu \, (u_\alpha F^{\alpha\beta} S_\beta)
   \nonumber \\
    & \qquad \qquad \qquad
    - \tau_R \, u^\mu 
				(S_\alpha F^{\alpha\beta} F_{\beta\lambda} u^\lambda) \,.\label{eqn:spincor}
\end{align}
The significance of the RR term in the second line of the T-BMT equation is to ensure orthogonality of the spin-vector and the velocity vector $S.u=0$ and the constancy of $S_\mu S^\mu$ for all times. It should be mentioned that these contributions are not small.
If neglected, the solutions of the  equations of motion vastly differ (see Figure~\ref{fig:SR}).

\subsection{Noncovariant form of the T-BMT equation: Rest-frame spin vector}\label{sect:BMT}

Here, we show that the RR term in the covariant T-BMT equation also gives a non-zero contribution to the precession of the spin-vector in the rest frame of the particle, and estimate the order of magnitude of the RR correction.

The equation of motion for
spin vector $\vec s$ in the electron rest-frame of the electron can be derived from the covariant BMT-equation and has the following form \cite{book:Jackson}:
\begin{align} \label{eq:BMT3}
    \deriv{\vec s}{\tau} = q \vec s \times \vec \Omega \,,
\end{align}
where
\begin{align} \label{eq:sS}
    S^\mu = \left( \vec u\cdot \vec s , \vec s + \frac{\vec u ( \vec u \cdot \vec s) }{1+\gamma} \right) \,, 
    \quad
    \vec s = \vec S  - \frac{\vec u ( \vec u \cdot \vec S)}{\gamma(1+\gamma)} 
\end{align}
The angular velocity vector $\vec \Omega$ around which $\vec s$ precesses is derived straightforwardly from the covariant BMT equation \eqref{eq:BMT-dotu}, which first yields
\begin{align} \label{eq:s_y}
    \deriv{\vec s}{\tau} 
    &= \frac{qg}{2} \left[ \vec y - \vec u \frac{y^0}{(1+\gamma)} \right]
    - \frac{\vec s \times (\vec u\times  \dot{\vec u} ) }{1+\gamma} 
    \,,
\end{align}
with $y^\mu = (y^0,\vec y) = F^{\mu\nu} S_\nu - u^\mu (u.F.S)$ and the last term being the Thomas precession \cite{book:Jackson}.

Without radiation reaction taken into account, $\dot {\vec u}$ is governed by the Lorentz force equation and the angular velocity vector is given by
\begin{multline} \label{eq:OmegaNORR}
    \vec\Omega = (1+a_e\gamma) \vec B 
    - \left( a_e \gamma + \frac{\gamma}{1+\gamma}\right) (\vec v \times \vec E ) 
    \\
    -\frac{a_e\gamma^2}{1+\gamma} \vec v (\vec v\cdot \vec B) \,,
\end{multline}
where $\vec v = \vec u/\gamma$.
Here it should be emphasized that all quantities defining $\vec \Omega$ are taken in the lab frame, but $\vec s$ is the rest-frame spin vector.

With radiation reaction taken into account, we find that $\vec \Omega$ acquires a contribution in addition to Eq.~\eqref{eq:OmegaNORR}, due to the radiation reaction terms in the orbital equation Eqn.~\eqref{eq:RR}:
\begin{align} \label{eq:Omega_rev}
 \vec\Omega  \to  \vec\Omega + \delta \vec \Omega_\mathrm{RR} \,, \qquad \delta \vec \Omega_\mathrm{RR}
=  \frac{\tau_{R}}{q} \frac{\gamma}{1+\gamma} 
(\vec v \times \vec R ) \,,
\end{align}
where $\vec R$ is the spatial part of $R^\mu$. Clearly, the specific form of the radiation reaction correction to spin precession depends on the specific form of $\vec R$ in the various radiation reaction models. Taking here the most simple Herrera form, i.e. Landau-Lifshitz with negligible field gradients,
$\vec R_H = \vec L\times \vec B + \vec E(\vec E\cdot \vec u)$, with the Lorentz force vector $\vec L = \gamma \vec E + \vec u \times \vec B$ we find
\begin{multline}
\delta \vec \Omega_\mathrm{RR} = 
\frac{\tau_R \gamma^2}{q(1+\gamma)} \big[
     (\vec v \cdot \vec B)  (\vec E - \vec v \times \vec B)
     \\
     -
     (\vec v \cdot \vec E)  (\vec B - \vec v \times \vec E)
        \big] \,.
\end{multline}
Because of the cross product $ \vec v \times \vec R_H$ only the sub-leading terms in the ultra-relativistic expansion of $\vec R_H$ contribute to 
$\delta \vec \Omega_\mathrm{RR}$; the leading term drops out.

We can now estimate under what conditions the correction $\delta \vec \Omega_\mathrm{RR}$ becomes relevant. For ultrarelativistic particles, $\gamma \gg 1$, the magnitude of $\delta \vec \Omega_\mathrm{RR}$ can be estimated as $ ||\delta \vec \Omega_\mathrm{RR}|| \sim  \tau_R \gamma F^2$, where $F\sim ||\vec E||, ||\vec B||$ stands for the magnitude of the electromagnetic field {in the laboratory frame}.
We have to compare it with the typical magnitude of the usual precession vector $\vec \Omega$, Eqn.~\eqref{eq:OmegaNORR}.
The latter has two contributions which scale as $||\vec \Omega||\sim F$ and $ ||\vec \Omega|| \sim a_e\gamma F$ for the normal and anomalous precession, respectively. From this we can derive two criteria, a classical one and a quantum one, under which the contribution $\delta\vec \Omega_\mathrm{RR}$ becomes relevant.

First, the classical criterion compares the RR term with the normal precession. In order for the RR correction to become comparable we have to fulfill the criterion $\tau_R \gamma F \sim 1$. With $\tau_R\sim\alpha$ this can be rewritten as $\gamma F/F_\mathrm{class,cr} \sim 1$, where $F_\mathrm{class,cr}=F_S/\alpha$ is the classical critical field, with the Schwinger field $F_S$ \footnote{Recall that we had normalized the field strength to the Schwinger field strength $m^2/|e|$. Thus, formally $F_S=1$, which we write out explicitly here for clarity.}. Thus, according to this criterion the RR corrections to the BMT equation become relevant if the background field in the rest frame of the particle is on the order of the classical critical field, which is much larger than the critical field of QED, hence quantum effects can be expected to become important much earlier. This criterion can also be recast in the form $\alpha \chi \sim 1$, where $\chi$ is the quantum nonlinearity parameter (but $\alpha\chi$ is a classical parameter as the $\hbar$ in $\alpha$ and $\chi$ cancel).

The second criterion is a quantum one and related to the anomalous magnetic moment. It reads $\tau_R  F \sim a_e $. Taking for the anomalous moment the 1-loop Schwinger value, $a_e=\alpha/2\pi$, we obtain that the corrections become relevant if $F/F_S\sim 1$. Thus, according to this criterion radiation reaction effects for spin precession become relevant if the field in the laboratory frame is on the order of the Schwinger field.

Both of these corrections seem out of reach for present day (and near future) high-intensity laser-plasma or laser-beam interactions. Those are typically characterized by $F \ll F_S$ and $\chi \gtrsim 1$. However, $\alpha\chi \sim 1$, as required by the first criterion, is still hard to reach. Moreover, for both criteria the applicablity of the classical approach ceases to be valid for much weaker fields than required for RR to affect spin precession. In particular in the latter case one already approaches the regime where one might see a breakdown of the Furry expansion of strong-field QED (Ritus-Narozhny conjecture)\cite{fedotov_advances_2023}.

We thus can conclude that the kind of RR effects discussed so far are negligible from a phenomenological standpoint. Nonetheless, we should emphasize that RR effects are crucial in the covariant form of the T-BMT equation to ensure the orthogonality $u.S=0$ throughout the complete time-evolution.

This concludes our classical investigations of the interplay between radiation reaction effects and spin precession. In the following we will derive a kinetic description for polarized particles relevant for contemporary high-intensity laser-plasma interactions, taking into account quantum effects. From those we will then derive effective semiclassical single-particle equations of motion for radiating particles with spin.

\section{Spin dependent kinetic equations}\label{sect:kinetic}

Here, we derive Boltzmann-type kinetic equations for describing high-intensity laser interactions with polarized particles. In principle, quantum transport theory, e.g. the Wigner operator formalism \cite{Vasak:AnnPhys1987,bialynicki-birula_phase-space_1991} or the nonequilibrium 2PI approach~\cite{fauth_collisional_2021}, can provide equations for the evolution of statistical ensembles of particles to all orders in $\hbar$, and in the presence of strong background fields. However, the resulting equations are often extremely involved and thus require a number of approximations. 

{For high-intensity laser-plasma interactions it was found \cite{Gonoskov:PRE2015,Gonoskov_RMP_2022} that typically the electromagnetic spectrum consists of two well-separated regions: (1) A coherent low-frequency peak describing the background/external and plasma fields including coherent radiation, and (2) an incoherent high-frequency peak. This suggests that for the interaction of the fermions with a strong, weakly varying electromagnetic field one can perform a mean field (Hartree-type) approximation), in which fluctuations of the electromagnetic fields are neglected} {to lowest order and then introduced as a perturbation.} 

The quantum effects at $O(\hbar)$ in the interaction of the fermions with the strong mean-field background can typically be neglected if the field strength fulfils $F/\gamma F_S \ll1$,
i.e.~the fields are weaker than $\gamma$ times the Schwinger field, and the scale length of the background field $\ell=1/{||}\nabla_x \ln F{||}\gg \lambdabar_C $ is long compared to the Compton wavelength 
\cite{Gao:JMPA2020,fauth_collisional_2021}. Both of these criteria are usually extremely well fulfilled for high-intensity laser-plasma interactions. Some phenomena occurring when the field strength is on the order of the Schwinger field have been discussed recently in Refs.~\cite{al-naseri_kinetic_2020,brodin_plasma_2023}.

The dominant quantum effects in laser-plasma interactions are  hard photon emission by an electron and the decay of a photon into an electron positron pair. Those effects couple the fermionic degrees of freedom to the strong background field and (fluctuating) high-frequency field modes (photons), hence are beyond the mean field approximation \cite{fauth_collisional_2021}.

Our strategy, therefore, is as follows: We take the leading order of quantum transport theory following from the Wigner operator formalism to derive equations for the classical Vlasov-type transport of the particle number density and a spin density. Quantum effects will be included by constructing quantum collision operators with the spin-resolved photon emission (non-linear Compton) rates in the locally constant field approximation in their integral kernels. A rigorous derivation of the collisional quantum transport equations from first principles, but without special emphasis placed on the particle polarization, can be found for instance in a recent article by~\citet{fauth_collisional_2021}.

\subsection{Classical Advection}

Here, we derive the equations describing the classical transport of the on-shell particle density $f$ and the spin-density $a^\mu$ in a strong background field from projections of the Wigner operator in the mean field approximation. {Hence, we assume that the BBGKY hierarchy truncates at the one-body level, i.e. we ignore particle-particle correlations (lepton-lepton/lepton-ion collisions etc.). This is valid for relatively dilute, high energy particle distributions.} According to \citet{Vasak:AnnPhys1987}, the transport equations for the \emph{off-shell} {distributions} $\mathcal F(x^\nu,p^\nu)$ and $\mathcal A^\mu(x^\nu,p^\nu)$, which are the scalar and axial-vector Fierz components of the Wigner function,
read
\begin{align} \label{eq:F_transport}
    p\cdot {\mathcal D} \mathcal F &=0 \,, \\
    \label{eq:A_transport}
    p\cdot {\mathcal D} \mathcal A^\mu &=  - F^{\mu\nu} \mathcal A_\nu  \,,
\end{align}
to lowest order in $\hbar$, complemented by the subsidiary condition $p_\mu \mathcal A^\mu=0$, where $p^\nu$ is an off-shell momentum, $p^2\neq 1$.
At the same order, the operator ${\mathcal D}_\mu$ is given by
\begin{align}
{\mathcal D}_\mu = \frac{\partial}{\partial x^\mu} + F_{\mu\nu} \frac{\partial}{\partial p_\nu} 
    \,,
\end{align}
where $F_{\mu\nu}$ is the electromagnetic~mean field. {From quantum constraints at lowest order in $\hbar$ it follows that the distribution functions must be on-shell\cite{Vasak:AnnPhys1987}, which we implement by writing}
\begin{align} \label{eq:F_ansatz}
    \mathcal F(x^\nu , p^\nu ) = {2} \delta(p^2 - 1) \theta(p^0) f(t,\vec x, \vec p)\,, \\
    \label{eq:A_ansatz}
    \mathcal A^\mu(x^\nu , p^\nu ) = {2} \delta(p^2 - 1) \theta(p^0) a^\mu(t,\vec x, \vec p)\,,
\end{align}
which singles out the positive energy (electron) mass shell. The transport equations for the on-shell distributions follow straightforwardly by integrating Eqns.~\eqref{eq:F_transport} and \eqref{eq:A_transport} over $\ud p^0$. They read
\begin{align} 
\label{eq:f_classical}
\mathcal T f(t,\vec x,\vec p)&= 0\,,\\
\label{eq:a_classical}
\mathcal T a^\lambda(t,\vec x,\vec p)   &= q F^{\lambda \nu} a_\nu \,. 
\end{align}
From here on, the momentum $p$ will always be on-shell 
with energy $\epsilon_{\vec p}=\sqrt{1+\vec p^2}$ and $p^2=1$. 
We also explicitly introduced the sign of the particle charge $q$, i.e.~$q=-1$ for electrons. The corresponding transport equations for positrons can be obtained by setting $q=+1$.

The on-shell Vlasov-type transport operators on the left-hand side of Eqns.~\eqref{eq:f_classical} and \eqref{eq:a_classical} are
\begin{align} \label{eq:T}
 \mathcal T 
    = p.\partial_x - q p.F.\partial_p 
    = \epsilon_{\vec p} \partial_t + \vec p \cdot \nabla_{\vec x} + q \vec L\cdot \nabla_{\vec p} \,,
\end{align}
with the partial coordinate derivative $(\partial_x)_\mu = (\partial_t,\nabla_x)$, and the on-shell momentum partial derivative
$\partial_p = (0,\nabla_{\vec p})$, and
with $\vec L = \epsilon_{\vec p} \vec E + \vec p\times\vec B$.

Because of the subsidiariy condition $p_\mu \mathcal A^\mu=0$, the time-component of the axial vector is not independent but rather a function of the spatial components. {Therefore, similar to the treatment of the spin vector in} Section~\ref{sect:BMT}, we can introduce an axial-vector (spin-)density $\vec a$ in the rest-frame of the particle according to 
\begin{align}
a^\mu = \left(\vec p \cdot \vec a , \vec a + \frac{\vec p (\vec p\cdot \vec a)}{1+\epsilon_{\vec p}} \right)\;,
\end{align}
which fulfills the subsidiary condition $p.a=0$ automatically. {Working here with $\vec a$ instead of the covariant $a^\mu$ will simplify the discussion spin-dependent radiation effects due to the RR term in the covariant classical BMT equation \eqref{eqn:spincor}.}
The rest-frame spin-density $\vec a$ obeys the transport equation
\begin{align}
    \mathcal T {\vec a} = q  {\vec a} \times  \left[ \vec B - \frac{\vec p\times \vec E}{1+\epsilon_p}\right] \,.
\end{align}
We can recognize the term in the square brackets on the right-hand side of the equation as $\vec \Omega$, Eqn.~\eqref{eq:OmegaNORR}, but without the anomalous magnetic moment, $a_e=0$. Thus, so far the equations only describe the normal spin precession.

So far the {transport} equations are {equivalent to} purely classical {descriptions}, {and at this level do not mix $f$ and $\vec a$. A mixture comes about through quantum effects.} The quantum effects in the interaction of the electrons with the strong background fields have been neglected by going to the leading order in $\hbar$ in the quantum transport. The corresponding $\mathcal O(\hbar^1)$-terms have been calculated, e.g.~in Ref.~\cite{Gao:JMPA2020}, and contain for instance spin-gradient forces. But, as we argued above that they are negligible for the typical conditions we are interested in $(E/E_S\ll\epsilon_p, \ell \gg \lambdabar_C)$. See also Ref.~\citet{Thomas:2020} for a detailed estimate of the size of spin-gradient forces.

What is \emph{not} negligible, however, are the quantum effects due to a coupling of the charged particle dynamics in background fields to high-frequency photon modes. These interactions are the root cause of quantum radiation reaction and must be taken into account for a consistent description of high-intensity laser-plasma interactions. 
This will be done by adding to the right hand side of Eqns.~\eqref{eq:f_classical} and \eqref{eq:a_classical} appropriate collision operators. This approach is similar to what has been done previously in the literature \cite{Elkina:PRSTAB2011,Nerush:PhysPlas2011,Neitz:PRL2013,Bulanov:PRA2013}, where the kinetic equations were formulated for all particle polarization effects neglected. The details will be given below in Section \ref{sect:collision}.

As has been discussed recently in the literature\cite{Ilderton:PLB2013,Torgrimsson:NJP2021,heinzl_classical_2021}, for a consistent treatment of strong-field QED effects one needs to include also the effect of electron self-energy loop corrections at the same order in $\alpha$. As will be discussed in more detail in Section~\ref{sect:loop}, in the quantum kinetic approach the only effect of the loop is to provide the anomalous spin-precession.

The full transport equations for a polarized plasma will therefore be of the form
\begin{align} \label{eq:f}
\mathcal T f = \left( \frac{\partial f}{\partial \tau} \right)_\mathrm{rad} + \left( \frac{\partial f }{\partial \tau} \right)_\mathrm{loop}
\,,\\
\label{eq:a}
\mathcal T {\vec a} - q  {\vec a} \times \vec \Omega_0
= \left( \frac{\partial  {\vec a} }{\partial \tau} \right)_\mathrm{rad} + \left( \frac{\partial  {\vec a} }{\partial \tau} \right)_\mathrm{loop}  \,,
\end{align}
where $\vec \Omega_0$ denotes Eqn.~\eqref{eq:OmegaNORR} with $a_e=0$.

\subsection{Radiation Emission Contributions: Quantum Collision Operators}
\label{sect:collision}

    In this section we are discussing how to include the effect of radiation emission in Eqns.~\eqref{eq:f} and \eqref{eq:f}. While the transport operators $\mathcal T$ describe the interaction of the electrons with the strong background fields in the mean field approximation, the radiation emission involves a coupling to high-energy photons, i.e.~fluctuating field modes. To describe the photon emission process we will employ the Furry expansion of strong-field QED at $O(\alpha)$ \cite{fedotov_advances_2023}.

    {For ultrarelativistic particles $\epsilon_p\gg1$, interaction with a strong field, the formation length of the photons is short compared to the scale lengths of the strong background field, and the photon emission rate can be calculated in the locally constant crossed field approximation (LCFA). The requirements are that the normalized scalar ($E^2-B^2$) and pseusoscalar ($\vec E\cdot \vec B$) field invariants are small compared to both unity and the quantum nonlinearity parameter $\chi\sim \epsilon_p E$. The formation length of high-energy photons becomes short if the parameter $\xi = E \ell\gg 1$, with the field scale length $\ell $. For instance in a plane wave, we might identify $\ell = \lambdabar_L$ with the reduced laser wavelength $\lambdabar_L=1/\omega_L$ such that $\xi$ becomes the usual classical laser-intensity parameter. In the quantum regime
    for $\chi\gtrsim 1$ it was found that additionally one has to require $\xi^3/\chi \gg 1$ \cite{Dinu:PRL2016}. (See also Refs.~\cite{Blackburn2018,DiPiazza:PRA2018,ilderton_extended_2019} for further discussions of the limitations of the LCFA for soft photon emission.)}
    
    Under these conditions, photon emission is a \emph{local} process (a short-range interaction) and can be described by local collision operators, $\mathcal C_f = \left( \frac{\partial f }{\partial \tau} \right)_\mathrm{rad}$ and $\mathcal C_{\vec a} = \left( \frac{\partial  {\vec a} }{\partial \tau} \right)_\mathrm{rad}$, where
    each of the collision operators consists of the usual gain and loss terms, e.g.~$\mathcal C_f = \dot f_\mathrm{gain} - \dot f_\mathrm{loss}$.

    For ultrarelativistic particles, the photon is emitted into a narrow cone around its instantaneous velocity. We therefore can employ a collinear emission approximation, where the emitted photon momentum is assumed to be parallel to the electron momentum at the moment of emission. This assumption breaks the energy conservation during emission, but the energy lack is of order $1/\epsilon_p^2$, i.e.~negligibly small for $\epsilon_p\gg1$ \cite{Blackburn:PRA2020}.

    The photon emission rates for polarized electrons can be represented using the Stokes vectors $\vec n_{i,f}$ of the incident/final electrons, which describe the state and degree of polarization of a particle. For instance $\vec n_i^2=1$ means the incident particles are perfectly polarized and $\vec n_f^2=0$ means the final particles are completely unpolarized. In our kinetic approach, the role of the Stokes vectors is taken by the \emph{local} polarization degree $\vec s(t,\vec x,\vec p) \equiv  { \vec a}(t,\vec x,\vec p) / f(t,\vec x,\vec p)$ of a plasma element in phase space.

    The Stokes vectors, together with the LCFA building blocks for the differential rates $w_A$, where the label $A$ runs over the various polarization contributions, completely describe the photon emission for polarized particles. The LCFA building blocks and the explicit formulae for the definition of the collinear differential photon emission rates ${w}_A(\vec p, \vec k)$ and the total photon emission rates $W_A(\vec p)$ are given in Appendix~\ref{app:rates}.

    \subsubsection{The gain term}

    The `gain' of the distribution functions at momentum $\vec p$ due to radiation emission must come from photon emission processes with \emph{final} electron momentum $\vec p$, integrated over all possible initial states that lead to said final state. If a photon with momentum $\vec k$ is emitted, the initial electron momentum was $\vec p+\vec k$ by means of the collinear emission approximation, integrated over all photon momenta $\vec k$. The integral kernels thus must be a linear combination of the distribution functions $f$, $\vec a$ at $\vec p+\vec k$ and the differential photon emission rates $w_A(\vec p+\vec k,\vec k)$.

    The probability for photon emission with the initial and final electron polarization taken into account can be compactly expressed with help of the electron Stokes vectors and Müller matrices\cite{Torgrimsson:NJP2021} as $R = \frac{1}{2} N_i M N_f^T$, where the Müller matrix collates the LCFA building blocks and $N_{i,f}=(1,\vec n_{i,f})$, see Appendix \ref{app:rates}. Calculating just the expression $N_i M$, with the identification of $\vec n_i\sim \vec s$, tell us the correct way how the LCFA building blocks have to be combined with the densities $f$ and $\vec a$:
    \begin{align}
        f N_iM = f(1,\vec s) M \sim 
        f \left(
        \begin{matrix}
        w_0 + \vec s \cdot \vec w_i \\ 
        \vec  w_f + \vec s \cdot \underline{w}_{if}
        \end{matrix}
        \right)        
        \sim { \dot f_\mathrm{gain} \choose \dot {\vec a}_\mathrm{gain}} 
    \end{align}

    Focusing for now on the gain term for $f$,
    \begin{multline}
        \dot f_\mathrm{gain}(\vec p) = 
        \int \frac{\ud^3\vec k}{\omega_k} \mathcal N f(\vec p+\vec k) [w_0(\vec p+\vec k,\vec k) 
        \\ 
        + \vec s(\vec p+\vec k) \cdot \vec w_i(\vec p+\vec k,\vec k) ] \,,
    \end{multline}
    where $\omega_k$ is the photon frequency and the normalization factor $\mathcal N = \epsilon_{\vec p}/\epsilon_{\vec p+\vec k} $ {has to be included because the rates are expressed as probability per unit proper time for momentum $\vec p+\vec k$, while the final change on the left-hand-side of the equation \eqref{eq:F_transport} is with regard to momentum $\vec p$.} The integral for $\dot {\vec a}_\mathrm{gain}$ is constructed analogously.

    \subsubsection{The loss term}

    The loss terms describe the `loss' of particles from the momentum `mode' $\vec p$ due to radiation emission. Since any radiation emission alters the electrons' momentum state, the final electron polarization does not matter for the loss term and should be summed over. In order to find the form of the loss terms it is useful to assume the case of electrons circulating a constant B-field. In this case the direction $\uvb$ is parallel to $\vec B$ in the lab frame, i.e.~stationary. The particles will be polarized along this axis, and the polarization vector is non-precessing. 

    Let us first introduce the fractions of particles $f^\sigma$ with spin up ($\sigma=+1$) or down ($\sigma=-1$) as 
    \begin{align} \label{eq:fupdown}
    f^\sigma 
    = \frac{1}{2}(f + \sigma \, \uvb \cdot \vec  a)  = \frac{1}{2}(f + \sigma a) \,.
    \end{align}
    In this representation, the loss terms for the $f^\sigma(\vec p )$ due to photon emission, hence the electron losing momentum, can be straightforwardly written as
    \begin{align}
        \dot f_\mathrm{loss}^\sigma(\vec p) =  f^\sigma(\vec p) \sum_{\sigma'=\pm1}  W^{\sigma\sigma'}(\vec p) \,, 
    \end{align}
    i.e.~with the total emission rates summed over all possible spin-flip and non-flip transitions. Here, $W^{\sigma\sigma'} = \frac{1}{2} ( W_0 + \sigma \, W_i + \sigma'\,  W_f + \sigma\sigma'\,W_{1}  )$, thus $\sum_{\sigma'} W^{\sigma\sigma'} =  ( W_0 + \sigma W_i ) $. With Eqn.~\eqref{eq:fupdown} it is now straightforward to read off the corresponding expressions for $f$ and $a$ as
    \begin{align}
        \dot f_\mathrm{loss} &=  f  W_0 +  a   W_i  \,,\\
        \dot  a_\mathrm{loss} &=   a  W_0 + f  W_i \,.
    \end{align}
    This can now be generalized to arbitrary directions of $\vec a$ as
    \begin{align}
        \dot f_\mathrm{loss} &=  f  W_0 + \vec a \cdot  \vec  W_i  \,,\\
        \dot  {\vec a}_\mathrm{loss} &=   \vec a  W_0 + f \: \vec W_i \,.
    \end{align}

    Combining everything from the preceeding subsection we obtain as results for the collision operators
    \begin{widetext}
    \begin{align}
        \label{eq:collop_f}
        \mathcal C_f(\vec p) = \int \frac{\ud^3\vec k}{\omega_k} \frac{\epsilon_{\vec p}}{\epsilon_{\vec p+\vec k}}
        \left[ f(\vec p+\vec k)  {w}_0 (\vec p+\vec k, \vec k) +  \vec a(\vec p+\vec k) \cdot {\vec w}_i(\vec p+\vec k,\vec k) \right] - f(\vec p)  W_0(\vec p) - \vec a(\vec p) \cdot {\vec W}_i(\vec p ) \,, \\
        \label{eq:collop_a}
        \mathcal C_{\vec a}(\vec p) = \int \frac{\ud^3\vec k}{\omega_k} \frac{\epsilon_{\vec p}}{\epsilon_{\vec p+\vec k}}
        \left[ f(\vec p+\vec k)  {\vec w}_f (\vec p+\vec k, \vec k) +  {\vec a}(\vec p+\vec k) \cdot {\underline w}_{if}(\vec p+\vec k,\vec k) \right] - f(\vec p) \vec W_i(\vec p) - \vec a(\vec p)   W_0(\vec p ) \,.
    \end{align}
    \end{widetext}

    \subsection{Electron self-energy loop contributions and anomalous precession} \label{sect:loop}

    As has been discussed recently in the literature, the inclusion of loop contributions at the same order in $\alpha$ as the emission processes is necessary to maintain unitarity \cite{Ilderton:PLB2013,Torgrimsson:NJP2021,heinzl_classical_2021}. For high-energy photon emission at order $\alpha$ the corresponding loop contribution originates from the interference of the electron self-energy loop with the $\mathcal O(\alpha^0)$ part of electron propagation. As has been argued, in e.g.~\cite{Torgrimsson:NJP2021}, a convenient way of implementing this can be achieved by taking the combined Müller matrices for emission, $M$, and the loop $M^L$, $M +M^L$ especially in the resummation of quantum radiation reaction effects \cite{torgrimsson_resummation_2021}. We have to adapt this to our kinetic description.

    The basic strategy for constructing the loop contribution to the kinetic equations will be as follows: Calculate for the loop contribution, e.g.~$(\partial f/\partial\tau)_{\mathrm{loop}}$, the same combinations of the LCFA building blocks as for the emission operator, just with the corresponding expressions replaced by the loop expressions $R_A \to R_A^L$ (see Eqns.~\eqref{eq:loop1}--\eqref{eq:loop3}). The loop insertion does not change the particle momentum, and the expressions $R_A^L$ given in Appendix~\ref{app:rates} are already integrated over all photon momenta running around the loop. Thus, in the ``gain''-parts of $(\partial f/\partial\tau)_{\mathrm{loop}}$ and $(\partial \vec a/\partial\tau)_{\mathrm{loop}}$ we actually do not have any expressions which have to be integrated over photon momenta. 

    For the scalar density $f$ this means that the loop contribution must be zero, and indeed,
    \begin{align}
    \left( \frac{\partial f }{\partial \tau} \right)_\mathrm{loop} = f R_0^L + \vec a \cdot \vec R_i^L - 
        f R_0^L - \vec a \cdot \vec R_i^L = 0  \,.  
    \end{align}
    For the axial-vector density we obtain the nonzero result
    \begin{align}
        \left( \frac{\partial  {\vec a} }{\partial \tau} \right)_\mathrm{loop} &= f \vec R_f^L + \vec a \cdot \underline{R}_{if}^L - 
        f \vec R_i^L - \vec a  R_0^L\nonumber\\
        &= \alpha  \int_0^1 \ud\lambda \lambda \frac{\mathrm{Gi}(z)}{\sqrt{z}} \tilde{\vec a} \cdot (\uvk \uve -\uve \uvk) \,,
    \end{align}
    where the $\uve$, $\uvk$ and $\uvb$ are unit vectors relating to the instantaneous rest frame field components and form an approximately orthogonal basis; see Appendix \ref{app:rates}.
    This is nothing but the anomalous spin-precession in disguise. To see this clearly we first note that the integral over the Scorer function Gi yields the one-loop field-dependent value of the anomalous magnetic moment
    \begin{align} \label{eq:anomalous_field}
        a_e(\chi) = \frac{\alpha}{\chi}  \int_0^1 \ud\lambda \lambda \frac{\mathrm{Gi}(z)}{\sqrt{z}} \,,
    \end{align}
    which was derived by \citet{Ritus:JETP1970}. For weak fields, $\chi\to0$, it approaches Schwinger's 1-loop value $a_e(\chi\to0)=\alpha/2\pi$, and $a_e(\chi)$ decreases monotonically to zero as $\chi$ increases. We thus have for the loop contribution to the collision operator for the axial vector the following result: 
    \begin{multline}
    \left( \frac{\partial  {\vec a} }{\partial \tau} \right)_\mathrm{loop} 
    = \chi a_e(\chi) \: {\vec a} \cdot (\uvk \uve -\uve \uvk)  \\
    = - \chi a_e(\chi) \: {\vec a} \times \uvb 
    = - \vec a\times \vec \Omega_\mathrm{anomalous}(a_e(\chi)) \,,
    \end{multline}
    ignoring a term in $(\uve\cdot\uvb) \uve$, which should be negligible under the LCFA approximation and
    with the anomalous part of the spin-precession vector, Eqn.~\eqref{eq:OmegaNORR},  $\vec \Omega_\mathrm{anomalous} = \vec \Omega - \vec \Omega_0 = a_e \chi \uvb$.

    In summary, we have shown here that in the kinetic approach the loop contribution gives exactly the anomalous term of spin-precession, with the field-dependent anomalous moment $a_e(\chi)$. Combining this with the normal precession, which was obtained from the classical transport, it becomes clear that the axial-vector density $\vec a(t, \vec x, \vec p)$ at each point in phase space precesses around a local value of $\vec \Omega$ as given in Eq.~\eqref{eq:OmegaNORR}, but with the field-dependent anomalous moment $a_e\to a_e(\chi)$.

    \section{Moment hierarchy and relativistic fluid equations}\label{sect:moments}

    {An infinite chain of equations describing the transport of bulk plasma properties, including particle number densities, energy density, pressure, heat, etc., can be obtained by calculating  momentum moments of the kinetic equations  \cite{braginskii_transport_1965} and truncated by some choice of closure to yield fluid equations for the electron and positron species in the plasma.}

    Here, we derive {two-}fluid equations for {the electron and positron components of a relativistic plasma}~\cite{DeGroot:1980dk} with the particle spin-polarization taken into account {up to second order}. Our aim will be, in particular, to work out the effects of the spin-dependence in the collision operators and how they affect radiation reaction. The moment equations for a nonrelativistic electron gas with spin effects were discussed in Ref.~\citet{hurst_semiclassical_2014}. A relativistic transport equation for particles with spin was given in \cite{Ekman:PRE2017} using an extended phase space, see also Refs.~\cite{Zamanian2010,Brodin:PPCF2011,zamanian_extended_2010}, and \cite{hurst_semiclassical_2014} for the equivalence of the extended phase space with our method.

    In order to calculate the relativistic moment equations we have to integrate the two {kinetic} equations for $f$ and $\vec a$ for a generic moment of the kernel $\Psi = \Psi(p^\alpha,p^\beta, ...)$ and with the Lorentz invariant measure $\int \ud^3 \vec p /\epsilon_{p}$. Introducing the short-hand notation for the momentum space integrals $\langle  Y \rangle = \int \! \frac{\ud^3 \vec p}{\epsilon_p} \, f Y  $ with $f$ as the weight we find
    \begin{align} \label{eq:moment_general_f}
        \frac{\partial}{\partial x^\mu} \langle \Psi p^\mu \rangle
        & 
        = - q F_{\mu}^{\ \nu}  \left\langle  p^\mu \frac{\partial \Psi}{\partial p^\nu}   \right\rangle
        + \int \frac{\ud^3 \vec p}{\epsilon_p} \Psi \mathcal C_f \,, \\
        \frac{\partial}{\partial x^\mu} \langle   \Psi p^\mu \vec s \rangle
        & 
        = - q F_{\mu}^{\ \nu}  \left\langle \vec s p^\mu \frac{\partial \Psi }{\partial p^\nu} \right\rangle
        + q \langle \Psi \vec s \times \vec \Omega \rangle \nonumber \\
        & \qquad \qquad \qquad \qquad + \int \frac{\ud^3 \vec p}{\epsilon_p} \Psi  \mathcal C_{\vec a} 
         \,,
    \end{align}
    where in terms involving the electromagnetic fields we have integrated by parts and neglected the surface contributions. We recall the definition of polarization degree $\vec s \equiv  { \vec a} / f$.

    \subsubsection{Moments of the scalar distribution $f$}

    \paragraph{Zero order moment}

    For the zero order moment, i.e. with the kernel function $\Psi=1$, of the scalar transport equation for $f$, we find number density current conservation
    \begin{align}
        \frac{\partial}{\partial x^\mu} J^\mu
        =  \int \frac{\ud^3 \vec p}{\epsilon_p} \Psi \mathcal C_f = 0\,,
    \end{align}
    where the current {density is}
    \begin{align}\label{eq:current}
    J^\mu = \int \frac{\ud^3 \vec p}{\epsilon_p} p^\mu f = \langle p^\mu \rangle \,.
    \end{align}
    
    Clearly, the derivative in the {field advection term} in Eqn.~\eqref{eq:moment_general_f} vanishes since $\Psi$ is constant. The integral over the collision operator $\mathcal C_f$ vanishes because the gain and loss terms {must} exactly cancel when integrated over all electron momenta $\vec p$ to conserve {the electron number during photon emission}. To see this explicitly one may conveniently substitute the integration variable in the gain term $\vec p = \vec q - \vec k$, with $\ud^3 \vec p=\ud^3 \vec q$ to arrive at
    \begin{widetext}
    \begin{align} \label{eq:0f_helper}
    \int \frac{\ud^3 \vec p}{\epsilon_p} \mathcal C^\mathrm{gain}_f(\vec p) &= \int \frac{\ud^3\vec q}{\epsilon_{\vec q}} \int \frac{\ud^3\vec k}{\omega_k} 
        \left[ f(\vec q)  {w}_0 (\vec q, \vec k) +  \vec a(\vec q) \cdot {\vec w}_i(\vec q,\vec k) \right] \nonumber \\ 
    &= \int \frac{\ud^3\vec q}{\epsilon_q} f(\vec q) \int \frac{\ud^3\vec k}{\omega_k}   {w}_0 (\vec q, \vec k)
    +
    \int \frac{\ud^3\vec q}{\epsilon_q} \vec a(\vec q)\cdot  \int \frac{\ud^3\vec k}{\omega_k} 
     {\vec w}_i(\vec q,\vec k)     
        =  \int \frac{\ud^3 \vec q}{\epsilon_q} \mathcal C^\mathrm{loss}_f(\vec q)
    \end{align}
    \end{widetext}

    \paragraph{First order moment}

    Next, we discuss the first order moment of $f$, with $\Psi=p^\mu$. The left-hand-side of the equation can be rewritten as the divergence of the energy-momentum tensor of the plasma, $T^{\mu\nu} = \langle p^\mu p^\nu \rangle$. Evaluating the momentum derivatives in the field-advection term it takes the form of the Lorentz-force. {Combining with the collision terms}, we obtain
    \begin{align} \label{eq:moment_1_f}
        \partial_\mu T^{\mu\nu} = q F^{\nu\mu}J_\mu -\langle p^\nu (I_0 +\vec s\cdot \vec I_i) \rangle \,,
    \end{align}
    where $I_A = \int \ud \lambda \lambda \frac{\ud R_A}{\ud \lambda}$ are the first moments of the photon spectrum, i.e. the normalized radiated power.

    We now need to show that the integral over the collision operator $\mathcal C_f$ with $\Psi=p^\mu$ yields the last term in \eqref{eq:moment_1_f}. To see this,
    we first make the same substitution $\vec p = \vec q - \vec k$ as in \eqref{eq:0f_helper}.
    {Considering energy conservation, we write} $\epsilon_{\vec p} = \epsilon_{\vec q} - \omega_k + \delta$, where $\delta$ is the energy {error introduced by not including the momentum contribution from the background fields \cite{Seipt_PRL_2017}}. For ultrarelativistic particles and in the collinear emission approximation\cite{Ridgers:JCompPhys2014} we have $\delta = 1/2|\vec q| - 1/2|\vec p|\approx \frac{\lambda}{2\epsilon_p (1-\lambda)}$. The corresponding term of the integrated collision operator is by a factor of 
    $1/\epsilon_p^2\ll1$ smaller than the {leading order} term and can therefore be neglected. The {leading order} term reads
    \begin{multline}
    \int \frac{\ud^3 \vec p}{\epsilon_p} p^\mu \mathcal C_f(\vec p) 
    = - \int \frac{\ud^3 \vec p }{\epsilon_p} \int \frac{\ud^3\vec k}{\omega_k} k^\mu
        \left[ f(\vec p)  {w}_0 (\vec p, \vec k) 
        \right. \\
        \left. +  \vec a(\vec p) \cdot {\vec w}_i(\vec p,\vec k) \right]  \,.
    \end{multline}
    For the integral over the photon momentum we  {make use of the definition of the rate $w_A$, which is given in the Appendix,}  Eqn.~\eqref{eq:w_rate}. By means of the delta function, the photon four-momentum $k^\mu = (|\vec k|, \vec k) \to \lambda (|\vec p|,\vec p)\simeq  \lambda p^\mu$, and thus
    \begin{multline}
    \int \frac{\ud^3 \vec p}{\epsilon_p} p^\mu \mathcal C_f(\vec p) =
    - \int \frac{\ud^3 \vec p}{\epsilon_p} p^\mu 
    \left[ f(\vec p) I_0 + \vec a(\vec p) \cdot  \vec I_i \right]
    \\
    =  - \langle p^\mu   ( I_0 + \vec s(\vec p) \cdot  \vec I_i \rangle ,
    \end{multline}
    This {represents} the (momentum averaged) effect of quantum radiation reaction due to hard photon emission. In particular, the term $\vec s(\vec p) \cdot  \vec I_i \sim s_B I_i$  describes the spin-polarization dependence of radiative energy loss.

    \paragraph{Second-order moment}

    For the second-order moment with weight $\Psi = p^\nu p^\lambda$ we are led to the definition of the stress-flow tensor \cite{hazeltine_fluid_2002} $U^{\mu\nu\lambda} = \langle p^\mu p^\nu p^\lambda \rangle $, the trace of which equals the current $U^{\mu\nu}_\mu = J^\nu$, and which obeys the transport equation
    \begin{multline} \label{eq:moment_2_f}
        \partial_\mu U^{\mu\nu\lambda} =
        q F^{\nu \mu} T_{\mu}^\lambda +     q F^{\lambda \mu} T_{\mu}^\nu
    + \langle p^\nu p^\lambda (K_0 +\vec s\cdot \vec K_i) \rangle \\
    -2 \langle p^\nu p^\lambda (I_0 +\vec s\cdot \vec I_i) \rangle \,.
    \end{multline}
    The quantities $K_A = \int_0^1 \ud\lambda \lambda^2 \ud R_A/\ud\lambda$
    are the second moments of the photon emission spectrum, and describe the electron momentum spreading due to the stochasticicy of quantized photon emission. The term in the second line of Eqn.~\eqref{eq:moment_2_f} is the spin-dependent radiative cooling effect.

    \subsubsection{Moments of the axial-vector equation $\vec a$}
    \paragraph{Zero-order moment}

    Working out the zeroth moment, $\Psi=1$, of the axial vector equation we derive an equation for the spin current 
    \begin{align} \label{eq:moment_1_a}
        \vec J_s^\mu = \int \frac{\ud^3 \vec p}{\epsilon_p} \, p^\mu \vec a(\vec p) =\langle p^\mu \vec s\rangle \,.
    \end{align}
    
    As for the scalar current, the field advection term vanishes.
    However, the spin current is not conserved. It obeys the transport equation 
    \begin{align} \label{eq:moment_0_a}
        \partial_\mu \vec J_s^\mu 
        &= 
        q \langle \vec s \times  \vec \Omega \rangle    + \langle \vec W_f- \vec W_i + \vec s\cdot W_{if} - \vec s W_0 \rangle \,.
     \end{align}
    The first term on the right-hand-side is follows straightforwardly from spin-precession term in the transport equation Eqn.~\eqref{eq:a}. It causes a depolarization of the plasma if $\vec s(t,\vec x,\vec p)$ precesses at different frequency at either different points in space (i.e. different field strength) or for different momenta \cite{Mane:RPP2005,Thomas:2020}. The second term stems from integral over the collision operator $\mathcal C_{\vec a}$, where the calculation proceeds similar to the scalar case $\mathcal C_f$ above. This second term is responsible for the radiative polarization of the electrons{, including the Sokolov-Ternov \cite{book:Sokolov} or Baier-Katkov-Strakhovenko \cite{baier_radiational_1967,bauier_radiative_1972} radiative polarization effects.}

    With help of the expressions given in Appendix \ref{app:rates}, the combination of rates in the radiative polarization term can be reformulated as
    \begin{align}
        \vec W_f &- \vec W_i + \vec s\cdot \underline{W}_{if} - \vec s W_0 \nonumber \\
        & =  \uvb ( W_f - W_i )
        + (s_B \uvb + s_E\uve ) W_3 + s_K \uvk W_4
        \nonumber \\
        &= \alpha \int_0^1 \! \ud \lambda \: 
        \frac{\lambda^2}{1-\lambda}
        \left[
        \vec s \frac{\mathrm{Ai}'(z)}{z} 
        - \uvb \frac{\mathrm{Ai}(z)}{\sqrt{z}}
        \right. \nonumber \\ 
        & \qquad \left.
        - s_K \uvk \left( \mathrm{Ai}_1(z)  + \frac{\mathrm{Ai}'(z)}{z}\right)
        \right] \,,
    \end{align}
    where we have expanded the spin-polarization $\vec s$ in the three principal directions along the electric (magnetic) field  $\uve$ ($\uvb$) in the particle rest frame, and $\uvk=\uve\times\uvb$, according to
    $\vec s = s_B \uvb + s_E \uve + s_K\uvk$.

    \paragraph{Higher moments}

    With the weight functions $\Psi = p^\nu$ and $\Psi = p^\nu p^\lambda$ we define the spin energy-momentum tensor $\vec T_s^{\mu\nu} = \langle p^\mu p^\nu \vec s \rangle$ and spin stress-flow-tensor $\vec U^{\mu\nu\lambda}_{\vec s} = \langle p^\mu p^\nu p^\lambda \vec s \rangle $, respectively.
    They obey the transport equations
    \begin{align}
        \partial_\mu \vec T_s^{\mu\nu}
        & =  q F^{\nu\mu} (\vec{J_s})_\mu 
        +
        q \langle p^\nu \vec s \times  \vec \Omega \rangle    \nonumber \\
        & \qquad + \langle 
            p^\nu ( \vec W_f- \vec W_i + \vec s\cdot \underline{W}_{if} - \vec s W_0 )
           \rangle  \nonumber \\
        & \qquad -\langle p^\nu (
        \vec I_f +\vec s\cdot \underline  I_{if}) 
        \rangle
        \,, \\
        \partial_\mu \vec U_s^{\mu\nu\lambda}
         & =  q F^{\nu\mu} (\vec{T_s})_{\mu}^{\lambda} 
        +   q F^{\lambda\mu} (\vec{T_s})_{\mu}^{\nu} 
        +
        q \langle p^\nu p^\lambda \vec s \times  \vec \Omega \rangle 
        \nonumber  \\
        & \qquad + \langle 
            p^\nu p^\lambda ( \vec W_f- \vec W_i + \vec s\cdot \underline W_{if} - \vec s W_0 )
           \rangle \nonumber \\
        & \qquad -2 \langle p^\nu p^\lambda (
        \vec I_f +\vec s \cdot \underline  I_{if}) 
        \rangle \nonumber \\
        & \qquad + \langle p^\nu p^\lambda (
        \vec K_f +\vec s\cdot \underline K_{if}) 
        \rangle
        \,.  \label{eq:moment_2}
    \end{align}

    \section{Single particle equations of motion}\label{sect:singleparticle}

    From the moment hierarchy we can now derive effective semiclassical single-particle equations of motion for a {radiating lepton including the effects of its spin polarization. Note that the spin polarization is a classical quantity representing the expectation of the four components of the particle spin vector and so the single particle equations should be interpreted as the trajectory of an ``average'' lepton and not of an individual lepton}. To find those, we first note that the current density for a point particle may be written as \cite{barut}
    \begin{align}
        J^\mu(x) = \int \! \ud \tau \, u^\mu(\tau) \delta^{(4)} (x - x(\tau)) \,.
    \end{align}
    {By analogy with Eq.~\eqref{eq:current}, using the definition in Eq.~\eqref{eq:moment_1_a} we deduce that the appropriate form of the spin current density for a point particle should be
    \begin{align}
        J_s^\mu(x) = \int \! \ud \tau \, u^\mu(\tau)\vec{s}(\tau) \delta^{(4)} (x - x(\tau)) \,.
    \end{align}
    }

    {Hence, the scalar and axial vector distribution functions for a single point particle are}
    \begin{align}
       f(t,\vec x,\vec p) &= \int \ud \tau  \epsilon_{\vec p} \delta^{(3)} ( \vec p - \vec u(\tau)) \delta^{(4)} (x-x(\tau)) \,, \\
       \vec a(t,\vec x,\vec p) &= \int \ud \tau \vec s(\tau) \epsilon_{\vec p} \delta^{(3)} ( \vec p - \vec u(\tau)) \delta^{(4)} (x-x(\tau)) \,.
    \end{align}

    {From these definitions of the distribution functions,} the energy momentum tensor for a point particle is
    \begin{align}
    T^{\mu\nu} = \int \! \ud\tau \: u^\mu(\tau) u^\nu(\tau) \delta^4(x-x(\tau))\,.
    \end{align}
    {From this, we can derive the momentum conservation equation, starting with}
    \begin{multline}
    \partial_\mu T^{\mu\nu}  
     = \int \! \ud\tau \:   u^\nu(\tau) u^\mu(\tau) \partial_\mu \delta^4(x-x(\tau))
     = \\
     - \int \! \ud\tau \:   u^\nu(\tau) \frac{d}{d\tau} \delta^4(x-x(\tau))
     =  \int \! \ud\tau \:   \frac{\ud u^\nu(\tau)}{\ud\tau}  \delta^4(x-x(\tau))\,.
    \end{multline}
    Thus, from Eq.~\eqref{eq:moment_1_f} we find
    \begin{align} \label{eq:orbit_result}
        \frac{\ud u^\mu}{\ud \tau} = q F^{\mu \nu}(x(\tau)) u_\nu(\tau) - u^\mu(\tau) (I_0(\chi) + \vec s(\tau) \cdot \vec I_i(\chi))\,,
    \end{align}
    where $\chi = \chi(x(\tau),u(\tau))$ is the local value of the quantum parameter at the particle location. This quasi-classical equation for the particle acceleration contains the Lorentz force, and additional a spin-dependent radiation reaction term as the second term on the right. In the limit of unpolarized particles $\vec s=0$, our result agrees with the literature, e.g. Refs.~\cite{Ridgers:JPP2017,Niel:PRE2017,Elkina:PRSTAB2011}.

    The quasi-classical equation for the polarization vector evolution is obtained from Eqn.~\eqref{eq:moment_1_a} as
    \begin{align} \label{eq:spin-quasiclassical}
    \frac{\ud \vec s}{\ud \tau} = q \vec s(\tau) \times \vec \Omega(\tau) +
    \vec W_f - \vec W_i + \vec s\cdot (\underline{W}_{if} -W_0 \vec 1)\,,
    \end{align}
    where again the four-divergence is transformed into a proper time derivative. Eqn.~\eqref{eq:spin-quasiclassical} describes the spin precession, here with the anomalous magnetic moment being the field-depdendent value $a_e(\chi)$, Eq.~\eqref{eq:anomalous_field}. The additional terms describe the radiative polarization of the electrons, hence are the generalization of the Sokolov-Ternov effect. In the limit of weak fields, i.e. in the lowest order in $\chi$ this equation is known as the Baier-Katkov-Strakhovenko equation \cite{baier_radiational_1967,bauier_radiative_1972}. 
    Both these effects have been recently employed for simulating spin effects in laser-electron interactions, see e.g.~Refs.~\cite{guo_stochasticity_2020,
    tang_radiative_2021,Li_PRL_2020_helicity_transfer}.
    While a derivation of the precession part had been given recently in \cite{Ilderton2020,Torgrimsson:NJP2021}, no formal derivation of the full equation is known to us.

\section{Discussion} \label{sect:discussion}

\subsection{Comparison of the kinetic equation results with classical RR and the spin-dependent Gaunt factor}

By comparing the kinetic equation results for the radiative energy loss with the classical RR equations we can subsume all quantum effects into a spin-dependent Gaunt factor $\mathfrak g_{\vec s}$:
\begin{align}
    \langle p^\mu (I_0 + \vec s\cdot \vec I_i) \rangle =
    \langle p^\mu  \mathfrak g_{\vec s}(\chi) I_\mathrm{class} \rangle \,,
\end{align}
where $I_\mathrm{class} = \frac{2}{3} \alpha m \chi^2$ is the lowest order term in the asymptotic expansion of $I_0$ as $\chi\to0$. Thus,
\begin{multline} \label{eq:gfactor}
    \mathfrak g_{\vec s}(\chi) = -\frac{3}{2\chi^2} \int_0^1 \! \ud\lambda \: \lambda 
    \left(
    \Ai_1(z) + 2g\frac{\Ai'(z)}{z}  \right. \\
   \left. +  \lambda \frac{\Ai(z)}{\sqrt{z} } (\vec s\cdot \uvb)
    \right) \,.
\end{multline}
By using the tables in Appendix \ref{app:tables}, we can write down the series expansion of the spin-dependent Gaunt factor for small $\chi\ll1$ as
\begin{multline}
\mathfrak g_{\vec s}(\chi)
{\simeq} 
 1 
 - \left( \frac{55\sqrt{3}}{16 } + \frac{3}{2} s_B \right)\chi 
 + \left(48 + \frac{105 \sqrt{3} }{8} s_B \right) \chi^2 \\
 - \left( \frac{8855\sqrt{3}}{32 } + 300 s_B \right) \chi^3 \,.
\end{multline}

\begin{figure}
    \centering
    \includegraphics[width=0.99\columnwidth]{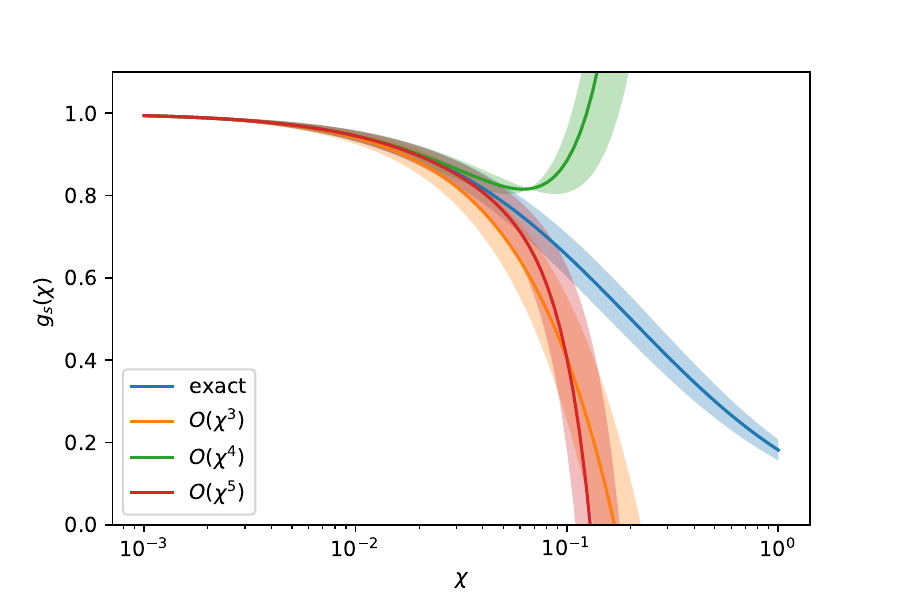}
    \includegraphics[width=0.99\columnwidth]{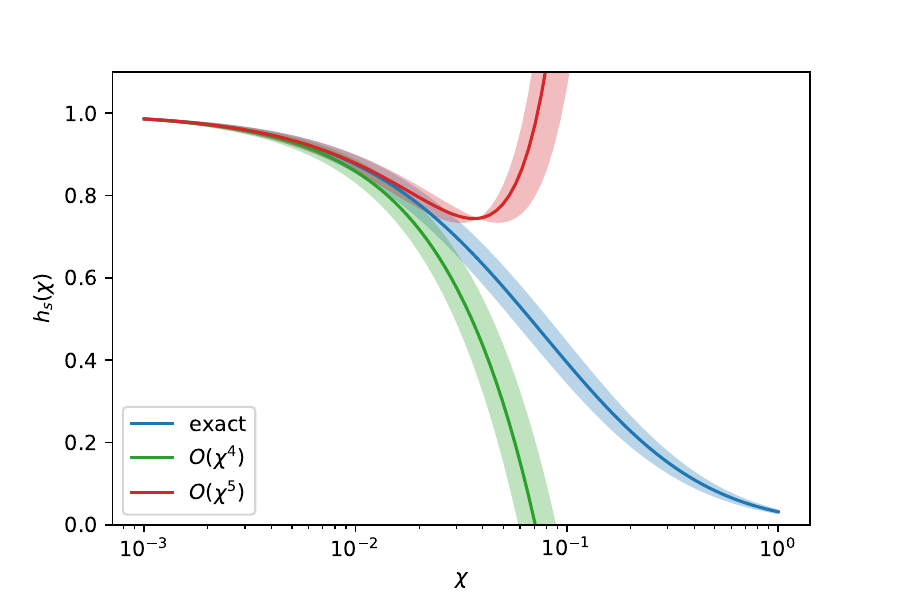}
    \caption{Gaunt factor $\mathfrak g_{\vec s}$ (upper) and analog quantity $\mathfrak h_{\vec s}$ for the spreading (lower). The bands indictate the strength of the spin effect, $-1\leq s_B\leq +1$. Curves have the following meaning: Exact (blue, Eqns.~\eqref{eq:gfactor} and \eqref{eq:hfactor} ) and asymptotics for small $\chi$ up to $\chi^3$ (orange), $\chi^4$ (green) and $\chi^5$ (red).}
    \label{fig:gfactor}
\end{figure}

As can be seen from this equation, only the degree of polarization along the rest frame magnetic field, $s_B = \vec s\cdot \uvb$, affects the Gaunt factor.
Positive values of $s_B$ reduce the Gaunt factor and hence radiative losses, while negative $s_B$ increase $\mathfrak g_{\vec s}$.
Thus, in an arbitrary electromagnetic field, electrons which have some degree of polarization in the direction of the rest frame magnetic field experience less radiation reaction than unpolarized electrons. The polarization orthogonal to $\uvb$ does not affect the strength of radiative losses. The function $\mathfrak g_s$ is plotted in the upper panel of Fig.~\ref{fig:gfactor}

By inspecting the effective single-particle equation, Eq.~\eqref{eq:orbit_result}, it is straightforward to find that \eqref{eq:orbit_result} violates the relativistic constraint $\dot u.u=0$\cite{Elkina:PRSTAB2011,Ridgers:JPP2017,Niel:PRE2017}. This behavior is a consequence of the ultra-relativistic and collinear emission approximation made in the collision operators, under which we kept only the leading order terms in inverse power of $\gamma$. It can be seen that in the limit $\mathfrak g_{\vec s}\to 1$ Eq.~\eqref{eq:orbit_result} agrees with the leading order ultrarelativistic approximation of the classical RR force in the Herrera form. This immediately indicates how to restore $\dot u.u=0$ for Eqn.~\eqref{eq:orbit_result}:
\begin{align}
\dot u^\mu = q F^{\mu\nu} u_\nu + \tau_R \mathfrak g_{\vec s} \Delta^{\mu}_\nu F^{\nu\lambda} F_{\lambda\kappa} u^\kappa \,.
\end{align}
Similarly, our Eqn.~\eqref{eq:spin-quasiclassical} doesn't contain the classical RR corrections to spin-precession of Eqn.~\eqref{eq:Omega_rev}. From Eqn.~\eqref{eq:Omega_rev} it is clear that only contribution to $\delta \vec {\Omega}_\mathrm{RR}$ is from the component of RR perpendicular to $\vec p$. The collinear emission assumption thus precludes any corrections of this kind. However, as we discussed in Section~\ref{sect:BMT} the effects of $\delta\vec \Omega_\mathrm{RR}$ are negligible for typical high-intensity laser-plasma and laser-beam interactions.

It would be interesting to investigate whether one automatically gets the `correct' classical limit if the angular distribution of photon emission is taken into account correctly in the collision operators instead of relying on the collinear emission approximation \cite{Blackburn:PRA2020,dai_effects_2023}.

By analogy to $\mathfrak g_s$ we can also define a `Gaunt factor' $\mathfrak h_s$ for the spreading effect as the ratio of $K_0+\vec s\cdot \vec K_i$ and its leading asymptotic term for $\chi\to0$, $K_0\sim \frac{\alpha m \chi^3 55}{24\sqrt{3}}$ as
\begin{align}
    \mathfrak h_{\vec s}(\chi) 
    &= -\frac{24\sqrt{3}}{55\chi^3} \int_0^1 \! \ud\lambda \: \lambda^2 
    \left(
    \Ai_1(z) + 2g\frac{\Ai'(z)}{z} 
    \right. \nonumber  \\
    & \qquad \qquad\qquad \qquad  \hfill \left.
    +  \lambda \frac{\Ai(z)}{\sqrt{z} } (\vec s\cdot \uvb)
    \right) \nonumber \\
    & {\simeq}  1 - \left(  \frac{448 \sqrt{3}}{55} + \frac{63}{22} s_B \right) \chi 
    \nonumber \\
    & \qquad\qquad\qquad \qquad  \hfill + \left(\frac{777}{4} 
    + \frac{480 \sqrt{3}}{11} s_B\right) \chi ^2 \,.
    \label{eq:hfactor}
\end{align}
The asymptotic expansion for small $\chi$ shows that the spreading is reduced if the electron have positive $s_B$, and increased if $s_B$ is negative. The function $\mathfrak h_s$ and its asymptotic approximation for small $\chi$ is plotted in the lower panel of Fig.~\ref{fig:gfactor}.

\begin{figure}
	\centering
	\includegraphics[width=0.99\columnwidth]{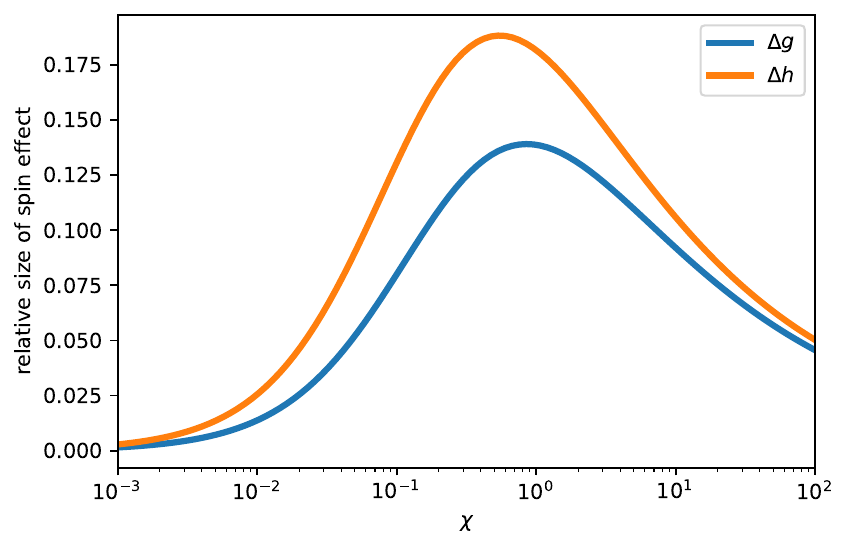}
	\caption{Relative size of the spin effect on the Gaunt factors $\mathfrak g_s$ and $\mathfrak h_s$.}
	\label{fig:relative_spin}
\end{figure}

To better quantify the relative size of the spin effect on $\mathfrak g_s$ and $\mathfrak h_s$, we plot in Figure~\ref{fig:relative_spin} the quantities $\Delta \mathfrak g \equiv (\mathfrak g_{-1} - \mathfrak g_{+1})/(\mathfrak g_{-1} + \mathfrak g_{+1})$ and $\Delta \mathfrak h \equiv (\mathfrak h_{-1} - \mathfrak h_{+1})/(\mathfrak h_{-1} + \mathfrak h_{+1})$. Both quantities vanish in the classical limit $\chi\ll 1$ and
peak at $\chi\sim 1$ with values of approximately 0.14 and 0.19, respectively. For even larger $\chi$ the relative spin contribution decreases.

\subsection{Evolution of average energy and average energy spread, and energy-polarization-correlation}

To define an average of a quantity such as the average energy of a plasma particle we need to divide the corresponding phase space integral by a suitable number density. {This choice is not unique and several definitions exist in the literature. To facilitate comparison of}  our spin-dependent results with spin-independent ones from the literature\cite{Ridgers:JPP2017}, we will {choose here to divide by the zero-component of the current four-vector $n=J^0$, corresponding to the laboratory frame density. This choice means that the equations lose their Lorentz covariance, but are more convenient to work with than other choices.}

We define a laboratory frame \emph{average} of a quantity $X$ as follows:
\begin{align}
    \bar{X} \equiv  \frac{\langle \epsilon_p X \rangle }{n} =  \frac{\langle \epsilon_p X \rangle }{\langle \epsilon_p \rangle} \,.
\end{align}
Thus, the average {fluid} \emph{velocity} is $\bar v^\mu = J^\mu /n = (1, \bar{\vec v})$,
where $\bar{\vec v} = \int \ud^3\vec p \, \vec v f / \int \ud^3\vec p f$ with $ {\vec v} = \vec p / {\epsilon_p}$.
We should note, however, that $\bar {v}^\mu$ is not a valid definition of a fluid \emph{four-velocity} as it does not square to unity; $\bar {v}_\mu \bar {v}^\mu = 1 - \bar {\vec v}^2 < 1$. But we can define a Lorentz factor, ${\bar{\gamma}} = (1 - \bar{\vec v}^2)^{-1/2}$ such that $u^\mu \equiv {\bar{\gamma}}  \bar {v}^\mu$ agrees with the {commonly used} Eckart frame fluid four-velocity \cite{DeGroot:1980dk} and $n = {\bar{\gamma}}  \sqrt{J.J}$,
see also Appendix~\ref{app:eckart}.

We can now define an average four-momentum as $\bar p^\mu = T^{0\mu}/n$,
average energy $\bar \epsilon_p = T^{00}/n$ {(equivalent to an ``effective temperature'' for the species)} and average spin {polarization} as $\bar{\vec s} = \vec J^0_{\vec s}/n = \langle \epsilon_p \vec s\rangle /\langle \epsilon_p\rangle$. Moreover, the average squared energy is $\overline{\epsilon_p^2} = U^{000}/n$.

With this, we first note that the current conservation yields the auxiliary equation
\begin{align}
    d_t n + n \nabla \cdot \bar{\vec v} 
    = 0\,,
\end{align}
having introduced the convective derivative $d_t =\partial_t +\bar{\vec v} \cdot \nabla $.

From Eqn.~\eqref{eq:moment_1_f} we can derive equations for the average energy and momentum. We'll focus here on the average energy, i.e.~$\nu=0$ component:
\begin{align}  \label{eq:energy_average}
n d_t \bar \epsilon_p
&=-\nabla \cdot \vec{Q} + q \vec{J}\cdot\vec E    
-{ \langle \epsilon_p ( I_0 + \vec s\cdot \vec I_i ) \rangle}
\end{align}
with $\delta\epsilon = \epsilon_p - \bar \epsilon_p$ {and 
$\vec{Q} = n \overline{\vec v \delta \epsilon }
=\langle\vec  p\delta \epsilon \rangle$
is the spatial part of the intrinsic heat flux four-vector $Q^\mu$, which is related to the total heat flux four-vector ${Q_T}^\mu = \langle\epsilon_p p^\mu\rangle$ through $Q^\mu = {Q_T}^\mu - \bar{\epsilon}_p J^\mu$}. The first term on the right-hand-side represents the  energy diffusion ({i.e. divergence of the} heat flux), the second term represents {Ohmic} energy gain, and the third term represents the radiative energy loss, {including the spin-dependence}.

Next we want to derive an equation for the evolution of variance of the energy, i.e.~the second central moment $\sigma^2_{\epsilon} = \overline{\epsilon_p^2} - {\bar \epsilon_p}^2$. From Eqn.~\eqref{eq:moment_2_f} we first derive the equation for $\overline{\epsilon_p^2}$
as
\begin{align}
nd_t \overline{\epsilon_p^2} &= 
    - \nabla \cdot \langle {\vec p} (\epsilon_p^2 - \overline{ \epsilon_p^2}) \rangle + q \vec{Q}\cdot\vec E 
    \nonumber \\
    & \qquad {+ \langle{ \epsilon_p^2(  K_0 + \vec s\cdot \vec K_i ) }\rangle 
     - 2 \langle{ \epsilon_p^2(  I_0 + \vec s\cdot \vec I_i ) }\rangle }\,.
\end{align}
With $d_t\sigma^2_{\epsilon} = d_t \overline{\epsilon_p^2} - 2 \bar \epsilon_p d_t \bar \epsilon_p$, and employing Eqn.~\eqref{eq:energy_average} we find for the energy variance
\begin{align}
    nd_t \sigma^2_{\epsilon} & = 
    - \nabla \cdot \langle{{\vec p} ( \delta \epsilon^2-\sigma_\epsilon^2)} \rangle
    +2 ( q \vec E  - \nabla \bar \epsilon_p ) \cdot \vec{Q}
    \nonumber \\
    & \qquad {+ \langle{\epsilon_p^2 (K_0 + \vec s\cdot \vec K_i)}\rangle
    - 2 \langle{\epsilon_p\delta\epsilon ( I_0 + \vec s\cdot \vec I_i) }\rangle} \,.
\end{align}
{The first term on the right hand side represents some measure of the transport of energy spread.} {With the interpretation of the average energy $\bar{\epsilon}_p$ as an effective temperature of the electrons,
the term $\nabla \bar{\epsilon}_p$ can be interpreted as thermoelectric {effective force. Analogous to the electric field acting on the current leading to energy gain/loss in Eqn.~\eqref{eq:energy_average}, the electric field and thermoelectric term act on the heat flux to increase the energy spread.}} The last two terms govern how the energy spreading of the plasma particles changes due to radiation effects including the particle polarization. According to Ref.~\cite{Ridgers:JPP2017}, where the corresponding terms were discussed for unpolarized electrons, the second-to-last term is the radiative heating of the beam due to the stochasticy of photon emission. The last term is a radiative beam cooling due to a phase space contraction because radiation losses are larger for higher-energy particles. 

For the evolution of average spin and the spin-energy-moment we find
\begin{align}
    n d_t \bar{\vec s} &= -   \nabla\cdot \Sigma
    + {q \left\langle \vec s \times {\vec \Omega }
    \right\rangle  }\nonumber \\
    & \qquad + 
    {\left\langle \vec W_f -\vec W_i + \vec s\cdot \underline{W}_{if} - \vec s W_0  \right\rangle 
     }\,, \\
    nd_t \overline{\epsilon_p \vec s} &=
    - { \nabla\cdot \left\langle\vec p (\epsilon_p\vec s - \overline{ \epsilon_p \vec s}) 
    \right\rangle }
    + q \vec{J}_s\cdot \vec E 
    + {q \left\langle \epsilon_p \vec s \times \vec \Omega  \right\rangle }\nonumber \\
    &\qquad + 
    {\langle \epsilon_p(\vec W_f -\vec W_i + \vec s\cdot \underline{W}_{if} - \vec s W_0 ) \rangle } \nonumber \\
    & {\qquad -  \langle \epsilon_p ( \vec I_f + \vec s\cdot \underline{I}_{if} ) \rangle } \,,
\end{align}
where we have defined the spin diffusion tensor $\Sigma = n\overline{\vec v \delta \vec s} = \langle \vec p \delta \vec s\rangle$, with $\delta \vec s = \vec s - \bar{\vec s}$.
This yields for the covariance between average spin-polarization and average energy,
$\mathrm{cov}(\epsilon_p, \vec s) = \overline{\epsilon_p \vec s} -  \bar \epsilon_p \bar{\vec s} $, the following evolution equation:
\begin{multline}
    n \, d_t \mathrm{cov}(\epsilon_p, \vec s) =
    - \nabla \cdot \left\langle \vec p (\delta \epsilon \delta \vec s - \mathrm{cov}(\epsilon_p,\vec s)) \right\rangle \\
     + ( q \vec E  - \nabla \bar{\epsilon}_p ) \cdot \Sigma  
     - \vec Q \cdot \nabla \bar {\vec s}
     \nonumber \\
      + q \langle \vec s\times\vec \Omega \delta\epsilon \rangle 
     + \langle (\vec W_f -\vec W_i + \vec s\cdot \underline{W}_{if} - \vec s W_0 ) \delta\epsilon \rangle \\
     - \langle \epsilon_p( \vec I_f + \vec s \cdot \underline{I}_{if} ) \rangle
     + \bar{\vec s} \, \langle \epsilon_p (I_0 + \vec s\cdot \vec I_i )\rangle
     \,.
\end{multline} 
In addition to a thermoelectric force driving the spin diffusion $\Sigma$, this equation also contains a force due to spin gradients $\nabla \bar{\vec s}$ coupling to the heat flux. The interpretation of the remaining terms is as follows: The terms in the third line stem from spin-precession and radiative polarization, respectively, and affect the correlation between spin and particle energy in a significant way if the particle energy spread $\delta \epsilon$ is large.
The two terms in the last line are coming from spin-dependent radiation reaction. If we denote the contributions from this last line by $n\delta_I \mathrm{cov}(\epsilon_p, \vec s)$ and
expand the spin-vector along the principal directions,
$\vec s = s_B\uvb +s_E\uve + s_k \uvk$, we obtain
\begin{align}
  \delta_I \mathrm{cov}(\epsilon_p, s_B)  
    &= \label{eq:cov_sB}
        - \overline{I_f} 
        - \overline{\delta s_B I_0}
        + \bar s_B \overline{s_B I_i}
        - \overline{s_B I_3} \,,\\
  \delta_I \mathrm{cov}(\epsilon_p, s_E)  
    &=  
        - \overline{\delta s_E I_0}
        + \bar s_E \overline{s_B I_i}
        - \overline{s_E I_3} \,, \\
  \delta_I \mathrm{cov}(\epsilon_p, s_K)  
    & = - \overline{\delta s_K I_0}
        + \bar s_K \overline{s_B I_i}
        - \overline{s_K I_4} \,,
\end{align}
which demonstrates the coupling of the different spin-polarization components; $s_B$ affects the correlations of $s_E$ and $s_K$, but not the other way around. In particular, for an initially unpolarized distribution only Eqn.~\eqref{eq:cov_sB} is nonzero.

\subsection{Sokolov-Ternov effect}

To discuss the Sokolov-Ternov effect from our kinetic equations, let us assume we have a globally constant direction $\uvb$ for all electrons. That could be for instance a uniform and homogeneous magnetic field $\vec B = B_0 \hat z$, with electrons circulating in a steady state with constant energy on cyclotron orbits, and with $p_z=0$. Sokolov and Ternov \cite{book:Sokolov} introduced the fractions $n^\sigma$ of electrons with spin-up ($\sigma=+1$) and spin-down ($\sigma=-1$) and formulated rate equations for those quantities involving the spin-flip rates.

Here, we can then define the number density of electrons with 
spin up/down, $n^\sigma$, analogous to Eqn.~\eqref{eq:fupdown} as 
\begin{align}n^\sigma = 
\frac{1}{2} \int 
    {\ud^3 \vec p}\:
    (f + \sigma a_z) 
= \frac{J^0 + \sigma \uvb \cdot \vec J_{\vec s}^0}{2}\,.
\end{align}

Since the integrated collision integral for $f$ vanishes, i.e. the current $J^\mu$ is conserved, as was shown above, the evolution equation for the $n^\sigma$ are essentially given by the momentum integral over $\mathcal C_a$ as in Eqn.~\eqref{eq:moment_0_a},
\begin{align}
  \left(  \frac{\partial n^\sigma}{\partial t}  \right)_\mathrm{rad}
  &=
  \frac{\sigma}{2} \int \! \frac{\ud^3\vec p}{\epsilon_p}\: f(\vec p)
   \left[   W_f - W_i   +   s_z W_3 \right] \,.
  \label{eq:ST}
\end{align}
This is not exactly identical to the textbook treatment of the Sokolov-Ternov effect, since on the right-hand-side of the equation it is not straightforward to identify the densities $n^\sigma$. Instead, the distributions $f$ and polarization degree $s_z$ are weighted with the photon emission rates.
In a homogeneous magnetic field we have $\chi= |\vec p| B_0 $ and thus different parts of the distribution function have different photon emission rates. That means, different parts of the distribution spin-polarize at different rate.

If we are looking for a stationary state of the polarization degree, then we should require that the integrand of on the right-hand-side of Eqn.~\eqref{eq:ST} must vanish for each value of $\vec p$ separately. This then leads to the Sokolov-Ternov equilibrium for each fraction of particles with parameter $\chi$. Using the asymptotic series of the rates for small $\chi$ in Appendix \ref{app:tables} we find
\begin{align}
    s_z^{eq}(\chi)  \sim 
    \frac{W_i-W_f}{W_3}
    \simeq
    - \frac{8}{5\sqrt{3}} + \frac{13}{50} \chi - \frac{92 \sqrt{3}}{125} \chi^2 \,.
\end{align}
The leading term is the textbook Sokolov-Ternov result \cite{book:Sokolov}. The $\chi$-dependent corrections predict that the equilibrium polarization degree decreases with increasing $\chi$. This phenomenon has been observed numerically for the radiative polarization of electrons in a rotating electric field Refs.~\cite{Sorbo_PRA,Sorbo_PPCF}.
Of course, the exact dynamics of spin polarization is more involved due to the temporal change of $\chi$ due to radiation reaction.

\section{Summary}\label{sect:conclusions} 

In this paper we have derived the kinetic equations for a spin-polarized relativistic plasma with the leading quantum effects due to the interaction of the leptons with a strong field taken into account via Boltzmann-type collision operators,
\begin{align}
\mathcal T f & = \mathcal C_f \,, \qquad 
\mathcal T {\vec a}  = q  {\vec a} \times \vec \Omega
+ \mathcal C_{\vec a}\,,
\end{align}
for the scalar density $f$ and the spin (axial-vector) density $\vec a$. The transport operator $\mathcal T$ is given in Eqn.~\eqref{eq:T}, the collision operators by Eqns.~\eqref{eq:collop_f}--\eqref{eq:collop_a}, and the precession vector $\vec \Omega$ by Eqn.~\eqref{eq:OmegaNORR}, with the field-dependent anomalous moment $a_e=a_e(\chi)$. The local quantum collision operators were derived using the locally constant field approximation of the photon emission rates and electron self-energy expressions. The loop contribution provides exactly the anomalous precession.

Starting from a classical description of  radiation reaction we discussed the relevance of radiation reaction for the covariant BMT equation and found a RR correction term to spin-precession. While this term is crucial for the evolution of the covariant spin-vector $S^\mu$, 
under typical conditions of high-intensity laser-plasma interactions this correction term is not relevant for the dynamics of the spin-vector $\vec s$ in the electron rest frame.

From the kinetic equations we have derived effective single-particle equations of motion for a spinning particle, taking into account the spin-dependent radiation reaction and radiative polarization.
We derived the moment hierarchy and obtained equations for the mean energy, energy spreading, and the energy-spin-correlation.

In this paper we focused on the interplay between the spin polarization and radiation reaction effects. Hence, the 
discussion was restricted to a polarized electron plasma without pair production effects taken into account. A generalization of our results to achieve a consistent description of a QED plasma of electrons, positrons and photons should be straightforward.

{
\acknowledgements
AGRT acknowledges support from the US NSF award \#2108075 and NSF/GACR award \#2206059.
The publication is funded by the Deutsche Forschungsgemeinschaft (DFG, German Research Foundation) -- 491382106, and by the Open Access Publishing Fund of GSI Helmholtzzentrum für Schwerionenforschung.
}

\appendix

\section{Polarization resolved photon emission rates}
\label{app:rates}

Here, we summarize all the polarization resolved photon emission rates appearing in the collision operators. Following Ref.~\cite{Elkina:PRSTAB2011} we define the differential rate for the collinear emission of a photon with momentum $\vec k$ by an electron with momentum $\vec p$ as
\begin{align} \label{eq:w_rate}
 w_{A}(\vec p,\vec k) = \int_0^1 \! \ud \lambda\, \omega_k \delta^{(3)}(\vec k-\lambda \vec p) \frac{\ud R_A}{\ud \lambda}\,,
\end{align}
where $\lambda$ is the fractional momentum transfer from the electron to the photon, and the label $A$ denotes the various polarization dependent contributions. The corresponding total rates are given by
\begin{align}
 W_A(\vec p ) = \int \frac{\ud^3 \vec k}{\omega_k}  w_{A}(\vec p,\vec k) = \int_0^1 \! \ud \lambda\,  \frac{\ud R_A}{\ud \lambda} = R_A\,,
\end{align}
where $\ud^3 \vec k/\omega_k$ is the Lorentz invariant phase space element of the emitted photon, with $\omega_k=|\vec k|$. The ${\ud  R_A}/{\ud \lambda}$ are the angularly-integrated LCFA rate building blocks for the polarization dependent photon emission rate (see e.g.~Ref.~\citet{Torgrimsson:NJP2021}), which are given here as a probability per unit proper time as
\begin{align}
\frac{\ud  R_0}{\ud\lambda} 
    & = - \alpha 
    \left[ \Ai_1(z) + 2g \frac{\Ai'(z)}{z} \right] \,, \\
\frac{\ud \vec  R_i}{\ud\lambda} 
    & = - \alpha 
     \lambda \frac{\Ai(z)}{\sqrt{z}} \: \uvb  
     =\frac{\ud  R_i}{\ud\lambda} \uvb \,, \\
\frac{\ud \vec R_f}{\ud\lambda} 
    & = - \alpha  
    \frac{\lambda}{1-\lambda} \frac{\Ai(z)}{\sqrt{z}} \: \uvb 
    =\frac{\ud  R_f}{\ud\lambda} \uvb \,, \\
\frac{\ud \underline R_{if}}{\ud\lambda} 
    & = 
            \frac{\ud R_1}{\ud\lambda}
                (\uvb \uvb +\uve\uve )
        + \frac{\ud R_2}{\ud\lambda}
                \uvk \uvk  \nonumber \\
   &    = \frac{\ud R_0}{\ud \lambda} \vec{1} 
              +      \frac{\ud R_3}{\ud\lambda}
                (\uvb \uvb +\uve\uve )
        + \frac{\ud R_4}{\ud\lambda}
                \uvk \uvk  
            \,, \\
\frac{\ud  R_1}{\ud\lambda} 
        &=  - \alpha 
            \left[  \Ai_1(z)+2\frac{\Ai'(z)}{z}   \right] \,, \\
\frac{\ud  R_2}{\ud\lambda} 
      & = - \alpha 
     \left[   (2g-1)\Ai_1(z) +2g \frac{\Ai'(z)}{z}  \right] \,,  \\
\frac{\ud  R_3}{\ud\lambda} 
        &= + \alpha  \frac{\lambda^2}{1-\lambda} \frac{\mathrm{Ai}'(z)}{z} \,, \\
\frac{\ud  R_4}{\ud\lambda} 
      & = - \alpha  \frac{\lambda^2}{1-\lambda} \mathrm{Ai}_1(z)\,.  
\end{align}
The label $A=0$ stands for the contribution of unpolarized particles, $A=i(f)$ is for initially(finally) polarized electrons. The different polarization correlation contributions ($A=if$) are be denoted by $A=1,2,3,4$. Ai is the Airy function with argument $z=(\lambda/\chi (1-\lambda))^{2/3}$, Ai$'$ its derivative, and Ai$_1$ the integral $\mathrm{Ai}_1(z) = \int_0^\infty \! \ud x \, \mathrm{Ai}(x+z)$, $g=1+\frac{\lambda^2}{2(1-\lambda)}$, and $\chi$ is the quantum parameter
\begin{multline} \label{eq:chi}
\chi = \sqrt{p.F^2.p} \\ 
= \sqrt{ (\epsilon_p \vec E(t,\vec x) + \vec p \times \vec B(t,\vec x) )^2 - (\vec p\cdot \vec E(t,\vec x))^2} \,.
\end{multline}

The principal directions $\uvb,\uve$ appearing in the above rates are unit vectors in the direction of the magnetic field and the electric field in the rest frame of the electron, and $\uvk = \uve \times \uvb$. In the ultrarelativistic approximation, $\epsilon_p \gg 1$, and when the angle between $\vec p$ and the field is not small we can write
\begin{align}
    \uve &= -q\frac{\vec E + \hat{\vec p} \times \vec B - \hat{\vec p}(\hat{\vec p}\cdot \vec E)}{\sqrt{(\hat{\vec p}\times \vec E)^2 + (\hat{\vec p}\times \vec B)^2 - 2\hat{\vec p}\cdot (\vec E\times\vec B)}} \,, \\
    \uvb &= -q\frac{\vec B - \hat{\vec p} \times \vec E - \hat{\vec p}(\hat{\vec p}\cdot \vec B)}{\sqrt{(\hat{\vec p}\times \vec E)^2 + (\hat{\vec p}\times \vec B)^2 - 2\hat{\vec p}\cdot (\vec E\times\vec B)}} \,,
\end{align}
where $\hat{\vec p} = \vec p / |\vec p|$. For positrons, with $q=+1$, the principal directions point opposite to the fields in the particle rest frame.
Under this approximation, in the rest frame of the particle the electromagnetic field appears to be a crossed field, thus $\uve \perp \uvb$, and $\uve$, $\uvb$, $\uvk$ form an orthonormal basis. Moreover, because of the collinear emission approximation these unit vectors are the same for incident and final particles.

The photon emission probability for polarized electrons can be compactly expressed employing the Müller matrix $M$ and the Stokes vectors $\vec n_{i,f}$ of the incident and final electrons as \cite{Torgrimsson:NJP2021,fedotov_advances_2023}
\begin{align}
   \frac{\ud R}{\ud \lambda} = \frac{1}{2} N_i M N_f^T \,,
\end{align}
where $N_{i,f} = (1,\vec n_{i,f})$. It should be noted that the $N_{i,f}$ are not Minkowski four-vectors, despite having four components.

The Müller matrix itself {is constructed by} collecting the LCFA building blocks above into a $4\times4$ dimensional matrix emission probability rate
\begin{align} \label{eq:muller-emit}
M = \left(
\begin{matrix}
    \ud R_0/\ud\lambda & \ud \mathbf R_f/\ud\lambda \\
    \ud \mathbf R_i/\ud\lambda & \ud \underline{R}_{if}/\ud\lambda
\end{matrix}
\right)\,.
\end{align}

Equivalent representations for the polarization resolved nonlinear Compton rates have been recently given in \cite{chen_electron_2022}, and a subset of these rates for electrons spin-polarized along the B-field direction was also given in \cite{Seipt_PRA_2020}. The rates can be also represented in the form of a density matrix \cite{Seipt_PRA_2018}.

In addition to the photon emission contribution at $O(\alpha)$ we also need the $O(\alpha)$ contribution from the one-loop electron self-energy.
More precisely, these contributions are coming from the interference of the one-loop self-energy at $\mathcal O(\alpha)$ with the `free' propagation of the electron. The expression are given by \cite{Torgrimsson:NJP2021}
\begin{align} \label{eq:loop1}
    R_0^L &= - R_0 \,, \\
    \vec R_i^L &= \vec R_f^L = \alpha  \int \ud\lambda \lambda \frac{\mathrm{Ai}(z)}{\sqrt{z}} \uvb \,, \\
    \underline{R}^L_{if} &= R_0^L \mathbf{1}_3 + \alpha  \int \ud\lambda \lambda \frac{\mathrm{Gi}(z)}{\sqrt{z}} (\uvk \uve -\uve \uvk) \,,
    \label{eq:loop3}
\end{align}
where $\mathrm{Gi}$ is the Scorer function \cite{scorer}. 
Airy and Scorer functions are the real and imaginary parts of the integral
\begin{align}
  \mathrm{Ai}(z) + i \mathrm{Gi}(z) = \frac{1}{\pi} \int_0^\infty \ud t \: 
  e^{ i\left(zt+ \frac{t^3}{3}\right) }  \,,
\end{align}
The Müller matrix $M^L$ for the loop is constructed analogously to Eqn.~\eqref{eq:muller-emit} as \cite{Torgrimsson:NJP2021}
\begin{align} \label{eq:muller-loop}
M^L = \left(
\begin{matrix}
     R_0^L & \mathbf R_f^L \\
    \mathbf R_i^L &  \underline{R}_{if}^L
\end{matrix}
\right)\,.\end{align}

\section{Asymptotics of the photon emission spectrum moments}
\label{app:tables}

The moments of the photon emission spectrum are defined as follows
\begin{align}
 W_A = \int_0^1 \ud \lambda \frac{\ud R_A}{\ud\lambda} \,, \\
 I_A = \int_0^1 \ud \lambda \lambda \frac{\ud R_A}{\ud\lambda} \,, \\
 K_A = \int_0^1 \ud \lambda \lambda^2 \frac{\ud  R_A}{\ud\lambda}\,,
\end{align}
where the subscript $A$ runs over the different building blocks of the spin-dependent photon emission rate.
For for small $\chi\to0$ the asymptotic expansions of the $Q \in  \{ W_A,I_A,K_A \}$ can be written as 
\begin{align}
Q = \alpha  \sum_{n=0}^\infty \beta_n^{\{Q\}} \chi^n \,,
\end{align}
where the coefficients $\beta_n^{\{Q\}}$ are presented in Table~\ref{tab:coeffs}.

\begin{table*}[]
    \centering
    \caption{Coefficients $\beta_n^{\{Q\}}$ of the asymptotic expansion as $\chi\to0$ for all moments up to order $\chi^5$.}
    \label{tab:coeffs}
    \begin{tabular}{ccccccc} \toprule
          & $1$ & $\chi$ & $\chi^2$ & $\chi^3$ & $\chi^4$ & $\chi^5$ \\ \midrule
      $n$    & $0$ & $1$ & $2$ & $3$ & $4$ & $5$ \\ \midrule
    $W_0$ & 0 & $\frac{5 }{2 \sqrt{3}}$
              & $-\frac{4}{3} $
              & $\frac{35 }{4 \sqrt{3}} $
              & $-\frac{92}{3}  $
              & $\frac{13475 }{32 \sqrt{3}} $ \\ 
    $W_1$ & 0 & $\frac{5 }{2 \sqrt{3}}$ 
              & $-\frac{4}{3} $
              & $\frac{55}{8 \sqrt{3}}$
              & $ -\frac{56}{3} $ 
              & $\frac{6545}{32 \sqrt{3}}$ \\
    $W_2$ & 0 & $\frac{5  }{2 \sqrt{3}}$
              & $-\frac{4}{3}$
              & $\frac{175}{24 \sqrt{3}}$
              & $-\frac{62}{3} $
              & $\frac{7469}{32 \sqrt{3}} $ \\  
    $W_3$     & $0$
              & $0$
              & $0$
              & $-\frac{5\sqrt{3}}{8} $
              & $12$ 
              & $- \frac{1155 \sqrt{3}}{16}$ \\
    $W_4$     & 0 & 0 & 0 
              & $- \frac{35}{24 \sqrt3}$
              & $10$
              & $- \frac{1001\sqrt3}{16}$ \\
    $W_i$  & 0 & 0
                & $-\frac{\sqrt{3}}{4}  $
                & $3 $
                & $\frac{-105 \sqrt{3} }{8} $
                & $ 200 $ \\  
    $W_f$  & 0 & 0 
                & $-\frac{\sqrt{3}}{4}$ 
                &$ 2 $
                & $\frac{-105 \sqrt{3}}{16} $ 
                & $80$ \\
              \bottomrule
    \end{tabular}
    \quad
    \begin{tabular}{ccccccc} \toprule
 & $1$ & $\chi$ & $\chi^2$ & $\chi^3$ & $\chi^4$ & $\chi^5$ \\ \midrule
      $n$    & $0$ & $1$ & $2$ & $3$ & $4$ & $5$ \\ \midrule
      $I_0$ & 0 & 0
    & $\frac{2 }{3}$ 
    & $-\frac{55 }{8 \sqrt{3}}$
    & $32$
    & $-\frac{8855 }{16 \sqrt{3}}$ \\
$I_1$ & 0 & 0
    & $\frac{2}{3}$
    & $-\frac{55}{8 \sqrt{3}}$ 
    & $28$
    & $-\frac{6545}{16 \sqrt{3}}$ \\
$I_2$ & 0&0
        &  $\frac{2 }{3}$
        & $-\frac{55 }{8 \sqrt{3}}$
        & $\frac{86}{3}$
        & $-\frac{6853 }{16 \sqrt{3}}$ \\
$I_3$ & 0 & 0 & 0& 0& $-4$ & $\frac{385 \sqrt3}{8}$ \\
$I_4$ & 0 & 0 & 0 & 0
        & $-\frac{10}{3}$
        & $\frac{1001}{8\sqrt 3}$ \\
$I_i$ & 0&0&0
        & $-1$
        & $\frac{35 \sqrt{3} }{4}$
        & $-200 $ \\
$I_f$ & 0&0&0
        & $-1$
        & $\frac{105 \sqrt{3}}{16}$
        & $-120 $ \\ \bottomrule
\end{tabular}
\quad
\begin{tabular}{ccccccc} \toprule
      & $1$ & $\chi$ & $\chi^2$ & $\chi^3$ & $\chi^4$ & $\chi^5$ \\ \midrule
      $n$    & $0$ & $1$ & $2$ & $3$ & $4$ & $5$ \\ \midrule
$K_0$ &  0&0&0
     & $\frac{55 }{24 \sqrt{3}}$
     & $-\frac{56}{3} $
     & $\frac{14245}{32 \sqrt{3}}$ \\
$K_1$ & 0&0&0
    & $\frac{55 }{24 \sqrt{3}}$
    & $-\frac{56}{3} $
    & $\frac{6545}{16 \sqrt{3}}$ \\
$K_2$ &0&0&0
    & $\frac{55}{24 \sqrt{3}}$
    & $-\frac{56}{3} $
    & $\frac{3311}{8 \sqrt{3}}$ \\
$K_3$ & 0& 0& 0& 0& 0&  $-\frac{385 \sqrt{3}}{32}$  \\
$K_4$ & 0 & 0& 0& 0& 0&  $-\frac{1001}{32\sqrt 3}$ \\
$K_i$ &0&0&0&0
    & $-\frac{35 \sqrt{3}}{16}   $
    & $100 $ \\
$K_f$ &0&0&0&0
    & $-\frac{35  \sqrt{3}}{16}   $
    & $80 $ \\ \bottomrule 
\end{tabular}
\end{table*}

\section{A note on the fluid velocity and relativistic hydrodynamics}
\label{app:eckart}

Here, we briefly discuss how one could properly define a fluid velocity. According to the literature, e.g.~Ref.~\cite{DeGroot:1980dk}, there are two {commonly used} options for defining the fluid velocity $u^\mu$:
(i) The Eckart frame via the vanishing of particle number/mass diffusion (i.e. via the particle number current), and (ii) the Landau frame via the vanishing of energy diffusion as the normalized eigenvector of $T^{\mu\nu}$.

The Eckart approach {defines a fluid four-velocity through} $J^\mu = N u^\mu$, with the fluid velocity $u^\mu$ ($u^2=1$) and $N$ is the \emph{invariant number density}, i.e.~the number density measured in the fluid rest frame.
Now, the particle momentum $p^\mu$ may be split into contributions parallel and perpendicular to $u^\mu$ as
\begin{align}
    p^\mu = \kappa_p u^\mu + \Delta^{\mu\nu} p_\nu \,.
\end{align}
with $ \Delta^{\mu\nu} \equiv  \eta^{\mu\nu} - u^\mu u^\nu $ and $\kappa_p \equiv u_\mu p^\mu$. 
With this we have
$\langle p^\mu \rangle = \langle \kappa_p \rangle  u^\mu 
+ \langle p^\mu - \kappa_p u^\mu \rangle $
i.e.~the vanishing of the second term means there is no particle diffusion current in the Eckart frame. 
 The number density is defined as $N=\langle \kappa_p \rangle = J^\mu u_\mu = \sqrt{J^\mu J_\mu}$. By the primary definition we can also use $u^\mu = J^\mu / \sqrt{J.J}$ to define the fluid velocity. Particle number current conservation $\partial.J=0$ yields the following equation: $\dot N + N\theta = 0$, where $\theta = \partial_\mu u^\mu$ is the expansion scalar (four-divergence of the fluid velocity) and $\dot N = (u^\mu \partial_\mu)N$ is the comoving or convective derivative.

The energy momentum tensor can be expanded into its irreducible tensor components 
\begin{align}
T^{\mu\nu} = \langle p^\mu p^\nu \rangle 
= e u^\mu u^\nu
+ u^\mu H^\nu + u^\nu H^\mu - p\Delta^{\mu\nu} +\pi^{\mu\nu}\,,
\end{align}
with energy density $e=\langle \kappa_p^2\rangle$, energy diffusion current (equivalent to heat flow in the Eckart description) $H^\mu = \langle \kappa_p \Delta^{\mu}_{\nu}p^\nu \rangle$, 
hydrostatic pressure
$p = - \frac{1}{3} \Delta_{\mu\nu} \langle p^\mu p^\nu\rangle$,
and shear-stress (or viscous pressure) tensor  
$\pi^{\mu\nu} = \Delta^{\mu\nu}_{\alpha\beta} \langle p^\alpha p^\beta\rangle$,
with the  double-symmetric, traceless rank-4 projection operator orthogonal to $u^\mu$, 
\begin{align}
\Delta^{\mu\nu}_{\alpha\beta} = \frac{1}{2}( \Delta^\mu_\alpha \Delta^\nu_\beta + \Delta^\mu_\beta \Delta^\nu_\alpha) - \frac{\Delta^{\mu\nu} \Delta_{\alpha\beta}}{\Delta^\lambda_\lambda}\,.
\end{align}
Moreover, $\pi^\mu_\mu = 0$, $u_\mu\pi^{\mu\nu}=0$ and $u_\mu H^\mu = 0$. Thus, the trace of the energy momentum tensor $T^\mu_\mu = e-3p = \langle p^\mu p_\mu \rangle = \langle 1\rangle = \int \frac{\ud^3 \vec p}{\epsilon_p} f$.

\section*{References}
%


\end{document}